\documentclass[journal]{IEEEtran}
\IEEEoverridecommandlockouts
\usepackage{setspace}               
\usepackage{stfloats}
\usepackage{cite}
\usepackage{amsmath,amssymb,amsfonts}
\usepackage{upgreek}
\usepackage{graphicx}
\usepackage{subfigure}
\usepackage{textcomp}
\usepackage{xcolor}
\usepackage{amsmath}
\usepackage[caption=false,font=normalsize,labelfont=sf,textfont=sf]{subfig}
\usepackage[linewidth=1pt]{mdframed}
\usepackage[mathscr]{eucal}
\usepackage{balance}
\usepackage{epstopdf}
\usepackage{cases}
\usepackage{xcolor}
\usepackage{bbding}
\usepackage{cleveref}
\usepackage{multicol}       
\usepackage{multirow}       
\usepackage{array}          
\usepackage{makecell}
\usepackage{flushend}
\usepackage{booktabs}
\usepackage{amsthm}
\usepackage{changebar}

\definecolor{crimson}{RGB}{192,0,0}         
\definecolor{navy}{RGB}{47,85,151}         
\usepackage{colortbl}
\makeatletter                               
\newif\if@restonecol
\makeatother

\makeatletter
\newif\if@restonecol
\makeatother
\usepackage[linesnumbered,ruled,vlined]{algorithm2e}
\usepackage{algpseudocode}
\theoremstyle{plain}
\newtheorem{thm}{Theorem}

\newtheorem{coro}{Corollary}

\theoremstyle{plain}
\newtheorem{rem}{Remark}

\begin{document}
\title{Mobile Cell-Free Massive MIMO with Multi-Agent Reinforcement Learning: A Scalable Framework}
\author{{Ziheng~Liu,~\IEEEmembership{Student Member,~IEEE}, Jiayi~Zhang,~\IEEEmembership{Senior Member,~IEEE}, Yiyang~Zhu,~\IEEEmembership{Student Member,~IEEE}, Enyu~Shi,~\IEEEmembership{Student Member,~IEEE}, and Bo~Ai,~\IEEEmembership{Fellow,~IEEE}}
\thanks{Z. Liu, J. Zhang, Y. Zhu, E. Shi, and B. Ai are with the School of Electronic and Information Engineering and also with the Frontiers Science Center for Smart High-Speed Railway System, Beijing Jiaotong University, Beijing 100044, China (e-mail: \{zihengliu, jiayizhang, yiyangzhu, 21111047, boai\}@bjtu.edu.cn).}}
\maketitle
\begin{abstract}
Cell-free massive multiple-input multiple-output (mMIMO) offers significant advantages in mobility scenarios, mainly due to the elimination of cell boundaries and strong macro diversity. In this paper, we examine the downlink performance of cell-free mMIMO systems equipped with mobile-APs utilizing the concept of unmanned aerial vehicles, where mobility and power control are jointly considered to effectively enhance coverage and suppress interference.
However, the high computational complexity, poor collaboration, limited scalability, and uneven reward distribution of conventional optimization schemes lead to serious performance degradation and instability. These factors complicate the provision of consistent and high-quality service across all user equipments in downlink cell-free mMIMO systems. Consequently, we propose a novel scalable framework enhanced by multi-agent reinforcement learning (MARL) to tackle these challenges.
The established framework incorporates a graph neural network (GNN)-aided communication mechanism to facilitate effective collaboration among agents, a permutation architecture to improve scalability, and a directional decoupling architecture to accurately distinguish contributions. In the numerical results, we present comparisons of different optimization schemes and network architectures, which reveal that the proposed scheme can effectively enhance system performance compared to conventional schemes due to the adoption of advanced technologies. In particular, appropriately compressing the observation space of agents is beneficial for achieving a better balance between performance and convergence.
\end{abstract}

\IEEEpeerreviewmaketitle
\section{Introduction}
The rapid advancement of wireless communication has sparked considerable research interest in the development of beyond fifth-generation (B5G) and sixth-generation (6G) networks. Among these, 6G networks are anticipated to exhibit vastly superior communication capabilities compared to the existing 5G networks.
With a 100-fold increase in peak data rates, achieving speeds of terabits per second (Tb/s), and a tenfold reduction in latency, 6G will enable ultra-fast transmission and near-instantaneous communication \cite{[2],[3]}.
Additionally, the stringent requirement of 99.99999\% end-to-end reliability guarantees ensures robust and uninterrupted connectivity, making 6G networks ideal for mission-critical applications such as remote surgery and autonomous driving.
To realize these remarkable communication capabilities, extensive research has been conducted on various promising technologies, such as cell-free massive multiple-input multiple-output (mMIMO) \cite{[1],[5],[28]} and artificial intelligence-aided communication networks \cite{[6],[7],[8]}.
Among these technologies, cell-free mMIMO, an evolutionary paradigm shift in MIMO technology \cite{[1],[5]}, exploits a large number of geographically randomly distributed access points (APs) that concurrently serve all user devices (UEs) adopting the same time-frequency resources, thereby enhancing communication effectiveness.
Compared with the conventional cellular mMIMO, the major transformation in cell-free mMIMO lies in the deployment of far more APs than UEs. This important shift disrupts the cell boundary mechanism, significantly aiding in the reduction of inter-cell interference, which is a major performance bottleneck in densely populated cellular networks \cite{[11],[38]}.
\begin{table*}[t]
\centering
\fontsize{8.5}{9.6}\selectfont
\caption{Comparison of Relevant Research with This Paper.}
\label{Paper_comparison}
\begin{tabular}{!{\vrule width1pt}  m{4.7 cm}<{\centering} !{\vrule width1pt} m{10 cm}<{\centering} !{\vrule width1pt} m{1.8 cm}<{\centering} !{\vrule width1pt}}
\Xhline{1pt}
\rowcolor{gray!30} \bf  Technological Challenges  & \bf Advanced Technologies Adopted in This Paper & \bf Insights \cr
\Xhline{1pt}
\bf Uneven Service Quality   & \textbf{Mobility:} Network Coverage Enhancement & Performance \cr\hline
\bf Serious Interference   & \textbf{Power Control:} Interference Suppression & Performance \cr\hline
\bf High Computational Complexity   & \textbf{MARL:} Distributed Computing and Parallelization Strategy Search & Complexity \cr\hline
\bf Poor Collaboration & \textbf{GNN-aided Communication Architecture:} Collaboration Enhancement & Performance \cr\hline
\bf Limited Scalability & \textbf{Permutation Architecture:} State Space Compression & Convergence \cr\hline
\bf Uneven Reward Distribution & \textbf{Directional Decoupling Architecture:} Downlink SE Value Partitioning; \textbf{Attention-based Mechanism:} Weighted Partitioning  & Performance \cr\hline
\Xhline{0.6pt}
\end{tabular}
\end{table*}

However, due to the limited transmission power of APs and expensive wired fronthaul, the coverage range provided by cell-free mMIMO is limited and mainly applied in hotspot areas \cite{[33]}. This results in UE not being able to achieve better service quality beyond the coverage range of AP, especially for UEs at the network edge or in shadow areas of mountain buildings. Therefore, providing UEs with better coverage to achieve uniform service quality is particularly important. Recently, unmanned aerial vehicles (UAVs) have been found to have enormous potential in assisting wireless communication due to their advantages in flexible maneuverability and enhanced communication capabilities \cite{[33],[35]}. Specifically, UAVs can be freely and flexibly deployed, making them a key enabler to assist various scenarios, such as providing temporary enhanced coverage for emergency areas.

Inspired by this, it is desired to integrate UAVs as mobile-APs into existing cell-free mMIMO systems to expand network coverage \cite{[34]}, especially in emergency rescue scenarios. However, the unpredictability of the location of interference sources makes it challenging to allocate reasonable mobility management strategies for all mobile-APs, resulting in most previous studies assuming that APs are stationary \cite{[13],[14],[15],[16]}. Fortunately, the novel multi-agent reinforcement learning (MARL) technology can achieve adaptive mobility decisions to effectively cope with complex dynamic environments through collaboration among agents \cite{[30]}. This provides a more favorable solution for mobility optimization, assisting all mobile-APs adjust their positions to effectively avoid conflicts with other interference sources and enhance coverage. However, existing MARL schemes fail to effectively capture the complex relationships among agents, often resulting in low collaboration capabilities and communication efficiency \cite{[19]}. Conversely, a promising solution to address this challenge has been proposed that introduces a graph neural network (GNN) into existing architectures. Specifically, GNN aggregates and processes messages from adjacent nodes to ensure that each node can utilize contextual information from neighborhoods, thereby achieving more robust and cohesive strategies \cite{[20],[21],[22]}. Therefore, integrating GNN into MARL helps to further enhance mobility management and network collaboration, especially in complex dynamic environments.

Moreover, in the realm of cell-free mMIMO systems, the implementation of power control technologies is essential for optimizing overall system performance, especially in suppressing interference and enhancing signal quality. Meanwhile, given the distributed nature of mobile-APs and the necessity of effective coordination in power allocation to further serve UE \cite{[12]}, the role of power control has become more important in cell-free mMIMO systems.
Recently, the development of advanced power control schemes has received significant attention from both the industry and academic sectors, enabling them to dynamically adapt to ever-changing network conditions \cite{[13],[14],[15],[16]}. For instance, numerous schemes have been proposed, such as signal processing-based schemes \cite{[13]}, heuristic-based schemes \cite{[14]}, deep neural network (DNN)-based schemes \cite{[15]}, and MARL-based schemes \cite{[16]}. Although conventional schemes can achieve high spectral efficiency (SE) performance, they are often insufficient in addressing the complexities introduced by the dense deployment of mobile-APs while fulfilling heterogeneous service quality requirements. Meanwhile, although DNN-based schemes excel in managing densely deployed large-scale networks, their heavy reliance on extensive data sets may hinder their practical applications.
By contrast, MARL \cite{[17],[18],[29]} represents a disruptive strategy that overcomes the limitations of conventional power control schemes in terms of computational complexity and data dependency.
In general, MARL inherently complements the distributed nature of cell-free mMIMO by allowing multiple agents (e.g., mobile-APs or UEs) to collaboratively learn and optimize their power control strategies. This enhances both scalability and adaptability without relying on a centralized central processing unit (CPU) \cite{[17]}.
In particular, this decentralized architecture allows agents to develop intricate strategies through mutual interactions, effectively handling interference and resource allocation, thereby optimizing overall network performance \cite{[31],[32]}.

On the other hand, conventional MARL-based schemes also encounter various challenges, such as poor sample efficiency and limited scalability \cite{[20]}. This is because the joint state-action space of the network grows exponentially with the increase in the number of agents, especially in large-scale dense networks.
This growth can lead to a phenomenon known as dimension explosion. Therefore, it becomes increasingly imperative to appropriately reduce the size of the joint state-action space to optimize existing MARL networks \cite{[23],[24]}. Besides, despite various efforts that have been devoted to the literature, all works about cell-free mMIMO with MARL made the overly simplifying and idealistic assumption of consistent reward contributions \cite{[16],[20],[28],[29]}. This indicates that the contribution of all downlink agents (e.g., mobile-APs) in the network training process is the same, i.e., the SE values calculated at each UE are evenly distributed among all mobile-APs. However, it has been proven that reward partitioning without actual contribution basis may have a significant impact on performance and convergence \cite{[20]}. As a remedy, a more practical credit allocation strategy can be adopted to address this challenge, which deploys additional supervisory networks to assist each agent achieve a reasonable allocation of global rewards based on their contributions \cite{[25],[26],[27]}.

Motivated by the above discussions, to fully unleash the performance of mobile cell-free mMIMO systems, we investigate a novel scalable framework to address the challenges encountered in the optimization process. The comparison between potential technological challenges in existing research and the advanced technologies adopted is summarized in Table \uppercase\expandafter{\romannumeral1} and the major contributions of this paper are listed as follows:
\begin{itemize}
\item We analyze a cell-free mMIMO system equipped with mobile-APs and derive the achievable downlink SE expression. Then, we investigate a joint optimization problem of mobility and power control to effectively enhance coverage and suppress interference.
\item We design a dynamic permutation network that includes permutation invariance (PI) and permutation equivariance (PE) to permute the order of entities and compress state space, thereby tackling the dimension explosion problem inherent in large-scale dense networks. Moreover, we consider a hypernetwork that generates an infinite number of candidate weight matrices to ensure that the allocated solution space is not constrained.
\item To address the issues of poor collaboration and misaligned reward contributions, we introduce a GNN-aided communication architecture to enhance collaboration, and propose a directional decoupling architecture that embeds an intrinsic reward function to provide different incentives for each mobile-AP's contribution, thus enabling effective reward partitioning. Moreover, an attention-based mechanism is proposed, which relies on the weights generated by the obtained states and actions to further improve the rationality of reward partitioning.
\end{itemize}
\begin{figure}[t]
\centering
    \includegraphics[scale=0.675]{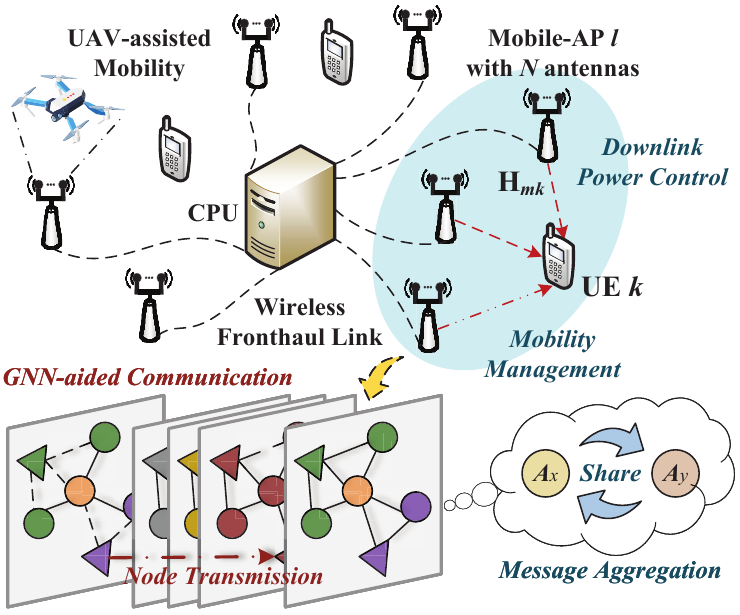}
    \caption{Illustration of a cell-free mMIMO system equipped with multi-antenna mobile-APs, which is embedded with a GNN-aided communication architecture to help all mobile-APs aggregate the observed partial information from neighboring mobile-APs and selectively combine received information.
    \label{fig1}}
\end{figure}

The rest of this paper is organized as follows. In Section \uppercase\expandafter{\romannumeral2}, we consider a cell-free mMIMO system and describe the corresponding channel estimation, downlink data transmission, and sum SE maximization. Next, Section \uppercase\expandafter{\romannumeral3} first introduces the implementation principles of permutation networks, then provides a detailed description of the modeling framework of dynamic hyper permutation networks. Then, in Section \uppercase\expandafter{\romannumeral4}, we propose a scalable permutation framework with MARL to address the challenges faced in joint optimization problems.
In Section \uppercase\expandafter{\romannumeral5}, numerical results, performance analysis, and a comparison of computational dimensions for the proposed scalable permutation framework with conventional MARL-based schemes are provided. Finally, the major conclusions and future directions are drawn in Section \uppercase\expandafter{\romannumeral6}.
\newcounter{mytempeqncnt_1}

\emph{\textbf{{Notation}}}: Boldface uppercase letters $\bf{X}$ and boldface lowercase letters $\bf{x}$ represent the matrices and column vectors, respectively. $\left(\cdot\right)$\textsuperscript{\emph{\textrm{T}}}, $\left(\cdot\right)$\textsuperscript{\emph{$\ast$}}, and $\left(\cdot\right)$\textsuperscript{\emph{H}} denote
the transpose, conjugate, and conjugate transpose, respectively. We utilize $\|\cdot\|$ and $|\cdot|$ to represent the determinant of a matrix and the Euclidean norm, respectively. $\mathbb{C}^n$ and $\mathbb{R}^n$ denote the $n$-dimensional spaces of complex and real numbers, respectively. $\nabla$, $\circ$, $\odot$ denote gradient, element-wise product, and module cascading, respectively. $\mathbb{E\{\cdot\}}$, $\text{tr}\{\cdot\}$, and $\triangleq$ are the expectation, trace, and definition, respectively.
Finally, $\mathbf{x} \sim {{\cal N}_\mathbb{C}}(\mathbf{0},\mathbf{R})$ represents the circularly symmetric complex Gaussian random variable with zero mean and correlation matrix $\mathbf{R}$.
\section{System Model}
In this paper, we consider a cell-free mMIMO system comprising $M$ mobile-APs equipped with $N$ antennas and $K$ single-antenna UEs, assuming that all mobile-APs simultaneously serve all UEs exploiting the same time-frequency resources, as illustrated in Fig. 1. Meanwhile, all mobile-APs achieve mobility optimization and downlink power control through mutual collaboration. Besides, the standard time-division duplex (TDD) protocol is adopted in our system with a coherence time block model consisting of $\tau_c$ time instants (channel uses), where pilot transmission occupies $\tau_p$ time instants and downlink data transmission consumes $\tau_c - \tau_p$ time instants.
Moreover, we denote the channel between mobile-AP $m$ and UE $k$ as $\mathbf{h}_{mk} \in \mathbb{C}^{N \times 1}$, $\forall m \in \{1,\ldots,M\}$, $\forall k \in \{1,\ldots,K\}$, which can be modeled through correlated Rayleigh fading and is given by
\begin{equation}
\setcounter{equation}{1}
\mathbf{h}_{mk} \sim {\cal N}_\mathbb{C}(\mathbf{0},\mathbf{R}_{mk}),
\label{eq1}
\end{equation}
where $\mathbf{R}_{mk} \in \mathbb{C}^{N \times N}$ denotes the spatial covariance matrix and $\beta_{mk} \triangleq \text{tr}(\mathbf{R}_{mk})/N$ represents the large-scale fading coefficient between mobile-AP $m$ and UE $k$.
\subsection{Channel Estimation}
For the channel estimation, we assume $\tau_p$ mutually orthogonal pilot sequences $\boldsymbol{\phi}_1,\ldots,\boldsymbol{\phi}_{\tau_p}$ are employed, which are assigned to all UEs in a deterministic but arbitrary manner, satisfying $K > \tau_p$ and $\|\boldsymbol{\phi}_t\|^2=\tau_p$, $\forall t \in \{1,\ldots,\tau_p\}$.
Besides, we denote the pilot sequence index assigned to UE $k$ as $t_k \in \{1,\ldots,\tau_p\}$ and represent the index subset of UEs adopting the same pilot sequence as UE $k$ (including itself) as $\mathcal{S}_k \subset \{1,\ldots,K\}$. Then, after all UEs send their pilot matrices in the uplink concurrently, the received signal $\mathbf{Y}_m^\text{p} \in \mathbb{C}^{N \times \tau_p}$ at mobile-AP $m$ can be modeled as
\begin{equation}
\setcounter{equation}{2}
\mathbf{Y}_m^\text{p} = \sum_{i=1}^{K}\sqrt{p_i}\mathbf{h}_{mi}\boldsymbol{\phi}_{t_i}^\text{T} + \mathbf{N}_m^\text{p}, \forall m,
\label{eq1}
\end{equation}
where $p_i \geqslant 0$ denotes the the transmit power of UE $i$, and $\mathbf{N}_m^\text{p} \in \mathbb{C}^{N \times \tau_p}$ represents the receiver noise with independent ${{\cal N}_\mathbb{C}}\left({0},\sigma^2\right)$ entries and noise power $\sigma^2$.

Firstly, to accurately derive sufficient statistical data for $\mathbf{h}_{mk}$ at mobile-AP $m$, we project $\mathbf{Y}_m^\text{p}$ onto $\boldsymbol{\phi}_{t_k}^*/\sqrt{\tau_p}$ as
\begin{equation}
\setcounter{equation}{3}
\begin{aligned}
\mathbf{y}_{mt_k}^\text{p}&= \frac{\mathbf{Y}_{m}^\text{p}\boldsymbol{\phi}_{t_k}^*}{\sqrt{\tau_p}} = \sum_{i=1}^{K}\frac{\sqrt{p_i}}{\sqrt{\tau_p}}\mathbf{h}_{mi}\boldsymbol{\phi}_{t_i}^\text{T}\boldsymbol{\phi}_{t_k}^* + \frac{1}{\sqrt{\tau_p}}\mathbf{N}_m^\text{p}\phi_{t_k}^*\\
&=\sum_{i \in \mathcal{S}_k}\sqrt{p_i\tau_p}\mathbf{h}_{mi}+\mathbf{n}_{mt_k},
\label{eq3}
\end{aligned}
\end{equation}
where $\mathbf{n}_{mt_k} \triangleq \mathbf{N}_m^\text{p}\phi_{t_k}^*/\sqrt{\tau_p} \sim {\cal N}_\mathbb{C}(\mathbf{0},\sigma^2\mathbf{I}_{N})$ represents the past-processing additive white Gaussian noise with variance $\sigma^2$.

Then, following the standard minimum mean square error (MMSE) approach in \cite{[5]}, the estimation $\hat{\mathbf{h}}_{mk}$ of the channel $\mathbf{h}_{mk}$ from mobile-AP $m$ to UE $k$ can be modeled as
\begin{equation}
\setcounter{equation}{4}
\hat{\mathbf{h}}_{mk} = \sqrt{p_k\tau_p}\mathbf{R}_{mk}\boldsymbol{\Psi}_{mt_k}^{-1}\mathbf{y}_{mt_k}^\text{p},
\label{eq4}
\end{equation}
where $\boldsymbol{\Psi}_{mt_k} \triangleq \mathbb{E}\{\mathbf{y}_{mt_k}^\text{p}(\mathbf{y}_{mt_k}^\text{p})^H\} = \sum_{i \in \mathcal{S}_k}p_i\tau_p\mathbf{R}_{mi} + \sigma^2\mathbf{I}_{N}$ denotes the correlation matrix of the received signal $\mathbf{y}_{mt_k}^\text{p}$. Note that the estimation error $\tilde{\mathbf{h}}_{mk} = \mathbf{{h}}_{mk}-\hat{\mathbf{h}}_{mk}$ and the channel estimation $\hat{\mathbf{h}}_{mk}$ are both independent random vectors, and their distributions can be represented as
\begin{equation}
\setcounter{equation}{5}
\tilde{\mathbf{h}}_{mk} \sim {\cal N}_\mathbb{C}(\mathbf{0},\mathbf{C}_{mk})
\label{eq5}
\end{equation}
with
\begin{equation}
\setcounter{equation}{6}
\begin{aligned}
\mathbf{C}_{mk} &= \mathbb{E}\{\tilde{\mathbf{h}}_{mk}\tilde{\mathbf{h}}_{mk}^H\} = \mathbf{R}_{mk} - \hat{\mathbf{R}}_{mk}\\
&=\mathbf{R}_{mk} - \sqrt{p_k\tau_p}\mathbf{R}_{mk}\boldsymbol{\Psi}_{mt_k}^{-1}\mathbf{R}_{mk},
\label{eq6}
\end{aligned}
\end{equation}
and
\begin{equation}
\setcounter{equation}{7}
\hat{\mathbf{h}}_{mk} \sim {\cal N}_\mathbb{C}(\mathbf{0},\sqrt{p_k\tau_p}\mathbf{R}_{mk}\boldsymbol{\Psi}_{mt_k}^{-1}\mathbf{R}_{mk}),
\label{eq7}
\end{equation}
where $\hat{\mathbf{h}}_{mk}$ and $\tilde{\mathbf{h}}_{mk}$ satisfy $\mathbb{E}\{\hat{\mathbf{h}}_{mk}^H\tilde{\mathbf{h}}_{mk}\} =0$.
\subsection{Downlink Data Transmission}
During the downlink data transmission, all antennas of all $M$ mobile-APs simultaneously serve all $K$ UEs via the same time-frequency resources, the received downlink signal ${y}_k$ at UE $k$ is given by
\begin{equation}
\setcounter{equation}{8}
\begin{aligned}
{y}_k &= \sum_{m=1}^{M}\sum_{i=1}^{K}\sqrt{\rho_{mi}}\mathbf{h}_{mk}^H\mathbf{w}_{mi}s_i + n_k,
\label{eq8}
\end{aligned}
\end{equation}
where $\rho_{mi} \geqslant 0$ denotes the the downlink power allocated to UE $i$ by mobile-AP $m$, and $\mathbf{w}_{mi} \in \mathbb{C}^{N \times 1}$ represents the normalized precoding vector, satisfying $\|\mathbf{w}_{mi}\|^2 = 1$ and $\mathbf{w}_{mi} = \tilde{\mathbf{w}}_{mi}/\|\tilde{\mathbf{w}}_{mi}\|$ with the normal precoding vector $\tilde{\mathbf{w}}_{mi}$. Moreover, $s_i \in \mathbb{C}^1$ and $n_k \sim {{\cal N}_\mathbb{C}}\left({0},\sigma^2\right)$ are the data symbol transmitted to UE $i$ with variance $\mathbb{E}\{|s_i|^2\} = 1$ and the independent receiver noise at UE $k$, respectively. Then, based on (8), we can derive the downlink achievable SE by adopting standard capacity lower bounds as the following corollary \cite{[5]}.
\begin{coro}
An achievable SE of UE $k$ in the downlink with the MMSE estimator is
\begin{equation}
\setcounter{equation}{9}
\mathrm{SE}_k = \left(1-\frac{\tau_p}{\tau_c}\right)\log_2\left(1+\mathrm{SINR}_k\right),
\end{equation}
where the effective SINR is given by
\begin{equation}
\setcounter{equation}{10}
\mathrm{SINR}_k = \frac{\Big(\boldsymbol{a}_k^T\boldsymbol{\mu}_k\Big)^2}
{\sum_{i=1}^{K}\boldsymbol{\mu}_i^T\mathbf{B}_{ki}\boldsymbol{\mu}_i - \Big(\boldsymbol{a}_k^T\boldsymbol{\mu}_k\Big)^2 + \sigma^2}
\end{equation}
with
\begin{align}
\boldsymbol{\mu}_k&=\big[\sqrt{\rho_{1k}};\ldots;\sqrt{\rho_{Mk}}\big],\\
\boldsymbol{a}_k&=\big[\mathbb{E}\{\mathbf{h}_{1k}^H\mathbf{w}_{1k}\};\ldots;\mathbb{E}\{\mathbf{h}_{Mk}^H\mathbf{w}_{Mk}\}\big],\\
b_{ki}^{mm'}&=\mathbb{E}\{\mathbf{h}_{mk}^H\mathbf{w}_{mi}\mathbf{w}_{m'i}^H\mathbf{h}_{m'k}\},\\
\mathbf{B}_{ki}&=\Big[b_{ki}^{11},\ldots,b_{ki}^{1M};\ldots;b_{ki}^{M1},\ldots,b_{ki}^{MM}\Big].
\end{align}
\end{coro}
\begin{IEEEproof}
The proof follows from the similar approach as \cite{[5]} and is therefore omitted.
\end{IEEEproof}

Note that Corollary 1 applies to any precoding scheme, and one promising choice is regularized zero-forcing (RZF) precoding, which can be represented as
\begin{equation}
\setcounter{equation}{15}
\tilde{\mathbf{w}}_{mk} = p_k\Big(\sum_{i=1}^{K}p_i\hat{\mathbf{h}}_{mi}\hat{\mathbf{h}}_{mi}^H + \sigma^2\mathbf{I}_N\Big)^{-1}\hat{\mathbf{h}}_{mk}.
\end{equation}
In contrast, considering another promising choice of precoding, e.g., maximum ratio (MR) transmission $\tilde{\mathbf{w}}_{mk} = \hat{\mathbf{h}}_{mk}$ that does not require any matrix inversion. In particular, its lower computational complexity makes it more suitable for cell-free mMIMO. As such, in the sequel, we apply MR precoding at all mobile-APs. Building upon this setting, we can obtain the closed-form SE expression in the following theorem.
\begin{thm}
For MR precoding $\tilde{\mathbf{w}}_{mk} = \hat{\mathbf{h}}_{mk}$, (9) can be derived in closed-form as
\begin{equation}
\setcounter{equation}{16}
\mathrm{SE}_{k,\mathrm{c}} = \left(1-\frac{\tau_p}{\tau_c}\right)\log_2\left(1+\mathrm{SINR}_{k,\mathrm{c}}\right),
\end{equation}
where the updated effective SINR is given by
\begin{equation}
\setcounter{equation}{17}
\mathrm{SINR}_{k,\mathrm{c}} = \frac{\Big(\boldsymbol{a}_{k,\mathrm{c}}^T\boldsymbol{\mu}_k\Big)^2}
{\sum_{i=1}^{K}\boldsymbol{\mu}_i^T\mathbf{B}_{ki,\mathrm{c}}\boldsymbol{\mu}_i - \Big(\boldsymbol{a}_{k,\mathrm{c}}^T\boldsymbol{\mu}_k\Big)^2 + \sigma^2}
\end{equation}
with
\begin{align}
\boldsymbol{a}_{k,\mathrm{c}}&=\Big[\mathbb{E}\{\mathbf{h}_{1k}^H\hat{\mathbf{h}}_{1k}\};\ldots;\mathbb{E}\{\mathbf{h}_{Mk}^H\hat{\mathbf{h}}_{Mk}\}\Big],\\
b_{ki,\mathrm{c}}^{mm'}&=\mathbb{E}\{\mathbf{h}_{mk}^H\hat{\mathbf{h}}_{mi}\hat{\mathbf{h}}_{m'i}^H\mathbf{h}_{m'k}\},\\
\mathbf{B}_{ki,\mathrm{c}}&=\Big[b_{ki,\mathrm{c}}^{11},\ldots,b_{ki,\mathrm{c}}^{1M};\ldots;b_{ki,\mathrm{c}}^{M1},\ldots,b_{ki,\mathrm{c}}^{MM}\Big].
\end{align}
\end{thm}
\begin{IEEEproof}
The proof follows from the similar approach as \cite{[5]} and is therefore omitted.
\end{IEEEproof}
\begin{figure*}[t]
\centering
    \includegraphics[scale=0.39]{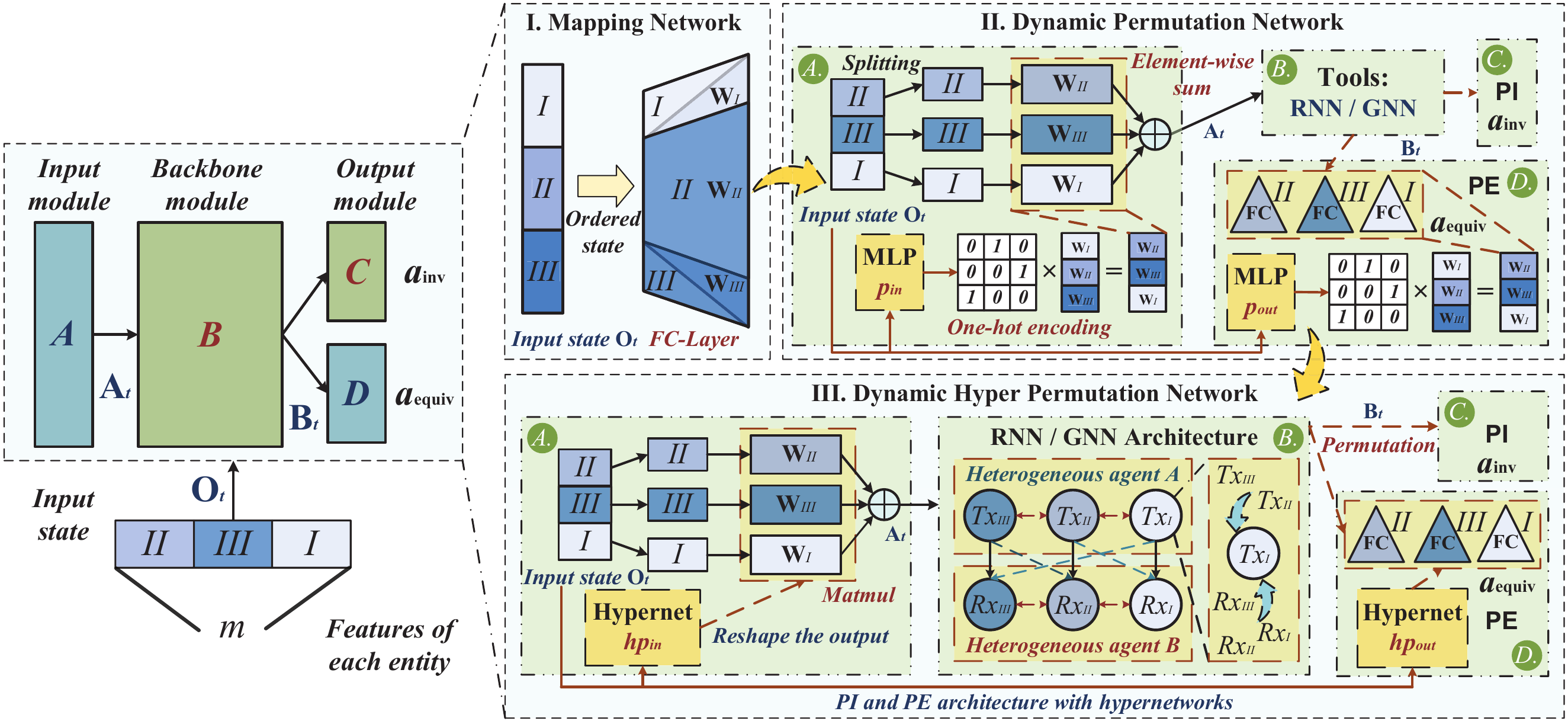}
    \caption{Illustration of the progressive evolution from a dynamic permutation network to a dynamic hyper permutation network, which consists of four parts: input module $\mathcal{A}$, backbone module $\mathcal{B}$, output module $\mathcal{C}$ for actions $a_\mathrm{inv}$, and output module $\mathcal{D}$ for actions $a_\mathrm{equiv}$.
    \label{fig1}}
\end{figure*}
\subsection{Sum SE Maximization}
To fully unleash the potential of cell-free mMIMO systems, it is crucial to study the problem of maximizing sum SE under mobility and power control constraints. By utilizing channel hardening characteristics, the normalized instantaneous channel gain can converge to the deterministic average channel gain.
This enables power $\rho_{mk}$ to be optimized based on large-scale fading coefficient $\beta_{mk}$ without considering fast time-varying small-scale fading, while $\beta_{mk}$ relies on the mobility of the mobile-AP to iteratively update.
Specifically, $\beta_{mk}$ is closely related to the position $(x_m,y_m)$ of mobile-AP $m$ and the position $(x_k,y_k)$ of UE $k$, which can be modeled by the large-scale fading function $\mathcal{L}(\cdot)$ as
\begin{equation}
\setcounter{equation}{21}
\beta_{mk}=\mathcal{L}(d_{mk})=\mathcal{L}(\sqrt{|x_m-x_k|^2+|y_m-y_k|^2}).
\end{equation}

Besides, since the constant prelog parameter $1-{\tau_p}/{\tau_c}$ in (9) does not affect the optimization process, we formulate the problem of maximizing the sum SE as
\begin{subequations}
\begin{align}
\max_{x_m,y_m,\rho_{mk}, \forall m,k} \sum_{k=1}^{K}\log&_{2}\left(1+\mathrm{SINR}_{k,\mathrm{c}}\right)\\
\mathrm{s.t.}  \qquad  \quad \thickspace \thinspace \thinspace d_{mk} &\geqslant d_{\min}, \forall m, k,\\
a_{m,\min} \leqslant  x_m,y_m &\leqslant a_{m,\max}, \forall m,\\
\sum_{k=1}^{K}\left\|\rho_{mk}\mathbf{w}_{mk}\right\|^2 &\leqslant P_{\mathrm{ap},\max}, \forall m,\\
|{w}_{mk}^{n}|^2 &\leqslant P_{\mathrm{an},\max}, \forall n,
\end{align}
\end{subequations}
where $d_{mk}$ and $d_{\min}$ denotes distance and minimum distance tolerance between mobile-AP $m$ and UE $k$, respectively. $a_{m,\min}$ and $a_{m,\max}$ represents the lower and upper limits of the movement area of mobile-AP $m$.
$P_{\mathrm{ap},\max}$ and $P_{\mathrm{an},\max}$ is maximum transmission power of each mobile-AP and antenna, respectively. Moreover, ${w}_{mk}^{n}$ represents the $n$-th element in the precoding vector $\mathbf{w}_{mk}$, $\forall n \in \{1,\ldots,N\}$.

It is worth emphasizing that the joint problem in (22) is non-convex, and conventional optimization schemes face various challenges, as shown in Table \uppercase\expandafter{\romannumeral1}, making them unsuitable for practical high-dimensional complex scenarios.
In the following sections, including the permutation model in Section \uppercase\expandafter{\romannumeral3} and the scalable framework in Section \uppercase\expandafter{\romannumeral4}, we develop a scalable permutation framework for practical cell-free mMIMO systems to address the aforementioned challenges.
\section{Permutation Model}
In this section, we first introduce the components and implementation principles of permutation networks, followed by a detailed description of the modeling framework for dynamic hyper permutation networks.
\subsection{Markov Decision Process Model}
In a typical multi-agent network, goal-oriented agents receive feedback by interacting with the environment, which is crucial for agents to learn and improve their decision-making abilities. In general, a common markov decision process (MDP) is utilized to model the sequential multi-agent decision-making process, which can be represented as a tuple $<\mathcal{O}_l, \mathcal{S}, \mathcal{A}_l, \mathcal{R}, \mathcal{T}, \gamma>$, where $\mathcal{O}_l$, $\mathcal{S}$, and $\mathcal{A}_l$ represent the observation space, state space, and action space of agent $l$, respectively \cite{[16]}. Specifically, each agent $l$ receives an observation $\mathbf{o}_{l,t} \in \mathcal{O}_l$ at time slot $t$, which contains partial information from global state information $\mathbf{s}_{l,t} \in \mathcal{S}$, and then obtains an action $\mathbf{a}_{l,t} \in \mathcal{A}_l$ based on its own policy $\pi_l(\mathbf{a}_{l,t}|\mathbf{o}_{l,t})$.
Moreover, $\mathcal{R}$, $\mathcal{T}$, and $\gamma$ denote the reward function, transfer function, and reward discount factor, respectively, satisfying $r_{\mathrm{ex},l,t} = \mathcal{R}(\mathbf{s}_{l,t},\mathbf{a}_{l,t})$ and $(\mathbf{s}_{l,t+1},\mathbf{o}_{l,t+1}) = \mathcal{T}(\mathbf{s}_{l,t},\mathbf{o}_{l,t+1},\mathbf{a}_{l,t})$.
\subsection{Permutation Policy Network: Methodology}
Due to the exponential growth of the dimension of the joint state-action space with the increase in the number of agents, many conventional MARL algorithms still suffer from challenges such as poor sample efficiency and limited scalability. By contrast, reducing the dimension of the joint state-action space by utilizing the inductive bias of PI and PE is a promising approach \cite{[24]}. Note that the permutation of inputs in the PI architecture does not change the function output, while the permutation of inputs in the PE architecture also permutes the outputs with the same permutation, i.e., the former satisfies
\begin{equation}
\setcounter{equation}{23}
\mathcal{F}\left(g[x_{l,t,1},\ldots,x_{l,t,f}]^T\right) = \mathcal{F}\left([x_{l,t,1},\ldots,x_{l,t,f}]^T\right),
\end{equation}
and the latter satisfies
\begin{equation}
\setcounter{equation}{24}
\mathcal{F}\left(g[x_{l,t,1},\ldots,x_{l,t,f}]^T\right) = g\mathcal{F}\left([x_{l,t,1},\ldots,x_{l,t,f}]^T\right),
\end{equation}
where $g \in \mathcal{G}_{\mathrm{p}}$ is a permutation matrix from a permutation set $\mathcal{G}_{\mathrm{p}}$, which ensures that there is a single unit value in each row and column, and zero elsewhere. $[x_{l,t,1},\ldots,x_{l,t,f}]$ denotes a dimension subdivision of $\mathbf{o}_{l,t}$ with $f$ features.

Moreover, we can divide all actions into two types based on the PI and PE architectures, including a type of entity-uncorrelated actions $\mathcal{A}_{l,\mathrm{inv}}$ and a type of entity-correlated actions $\mathcal{A}_{l,\mathrm{equiv}}$, satisfying $\mathcal{A}_{l}\triangleq(\mathcal{A}_{l,\mathrm{inv}},\mathcal{A}_{l,\mathrm{equiv}})$, where $\mathbf{a}_{l,t,\mathrm{inv}} \in \mathcal{A}_{l,\mathrm{inv}}$ should be invariant with the permutation of $\mathbf{o}_{l,t}$ and $\mathbf{a}_{l,t,\mathrm{equiv}} \in \mathcal{A}_{l,\mathrm{equiv}}$ should be equivariant. Specifically, given the same observation $\mathbf{o}_{l,t}$ arranged in different orders, the Q-values $Q_l(\mathbf{a}_{l,t,\mathrm{inv}}|\mathbf{o}_{l,t})$ of $\mathbf{a}_{l,t,\mathrm{inv}}$ should be kept the same, thus we can adopt PI architecture to make them learn more efficiently. On the other hand, for $Q_l(\mathbf{a}_{l,t,\mathrm{equiv}}|\mathbf{o}_{l,t})$ of $\mathbf{a}_{l,t,\mathrm{equiv}}$, due to the one-to-one correspondence between feature $\mathbf{o}_{l,t}$ and action $\mathbf{a}_{l,t,\mathrm{equiv}}$ in $\mathcal{A}_{l,\mathrm{equiv}}$ of each agent $l$, the permutations of $\mathbf{o}_{l,t}$ should result in the same permutations of $\mathcal{A}_{l,\mathrm{equiv}}$, so we can adopt PE architecture to make them learn more efficiently. Then, all actions under the permutation matrix $g$ can be modeled as
\begin{equation}
\setcounter{equation}{25}
\begin{aligned}
\mathbf{a}_{l,t} = (\mathbf{a}_{l,t,\mathrm{inv}},\mathbf{a}_{l,t,\mathrm{equiv}}),g\mathbf{a}_{l,t} = (\mathbf{a}_{l,t,\mathrm{inv}},g\mathbf{a}_{l,t,\mathrm{equiv}}).
\end{aligned}
\end{equation}

Correspondingly, by injecting the inductive bias of PI and PE architectures into the design of permutation networks, we can divide the network into four modules, including input module $\mathcal{A}_\mathrm{p}$, backbone module $\mathcal{B}_\mathrm{p}$, output module $\mathcal{C}_\mathrm{p}$ for actions $\mathbf{a}_{l,t,\mathrm{inv}}$, and output module $\mathcal{D}_\mathrm{p}$ for actions $\mathbf{a}_{l,t,\mathrm{equiv}}$, as shown in Fig. 2. The core idea is to adhere to the minimum modification principle, which only modifies $\mathcal{A}_\mathrm{p}$, $\mathcal{C}_\mathrm{p}$, and $\mathcal{D}_\mathrm{p}$, while keeping the backbone $\mathcal{B}_\mathrm{p}$ unchanged. This enhances the model's representational capability while maintaining the characteristics of PI and PE.
For simplicity, we denote the four modules in the entire permutation policy network as
\begin{equation}
\setcounter{equation}{26}
\begin{split}
\left \{
\begin{array}{ll}
    \mathbf{A}_{t} = \mathcal{A}_\mathrm{p}(\mathbf{O}_{t}),\mathbf{B}_{t} = \mathcal{B}_\mathrm{p}(\mathbf{A}_{t}),\\
    \text{PI's output: } \pi_l(\mathbf{a}_{l,t,\mathrm{inv}}|\mathbf{o}_{l,t}) = \mathcal{C}_\mathrm{p}(\mathbf{B}_{t})_l,\\
    \text{PE's output: } \pi_l(\mathbf{a}_{l,t,\mathrm{equiv}}|\mathbf{o}_{l,t}) = \mathcal{D}_\mathrm{p}(\mathbf{B}_{t})_l,
\end{array}
\right.
\label{eq25}
\end{split}
\end{equation}
where $\mathbf{O}_{t}=[\mathbf{o}_{1,t},\ldots,\mathbf{o}_{L,t}]^T$ is the observation matrix with the number of agents $L$. $\mathbf{A}_{t}$ and $\mathbf{B}_{t}$ represent the output matrices of $\mathcal{A}_\mathrm{p}$ and $\mathcal{B}_\mathrm{p}$, respectively. Then, to comply with the minimum modification principle, we can modify $\mathcal{A}_\mathrm{p}$ to PI and $\mathcal{D}_\mathrm{p}$ to PE relative to the observation $\mathbf{o}_{l,t}$, while keeping $\mathcal{B}_\mathrm{p}$ and $\mathcal{C}_\mathrm{p}$ unchanged as the following theorem and corollary.
\begin{thm}
If $\mathcal{A}_\mathrm{p}$ is made into PI, $\mathcal{C}_\mathrm{p}$ will immediately become PI without modifying $\mathcal{B}_\mathrm{p}$ and $\mathcal{C}_\mathrm{p}$, i.e.,
\begin{equation}
\setcounter{equation}{27}
\mathcal{C}_\mathrm{p}(g\mathbf{o}_{l,t}) = \mathcal{C}_\mathrm{p}(\mathbf{o}_{l,t}).
\end{equation}
\end{thm}
\begin{IEEEproof}
For the given observation $\mathbf{o}_{l,t}$, $\mathcal{A}_\mathrm{p}(g\mathbf{o}_{l,t}) = \mathcal{A}_\mathrm{p}(\mathbf{o}_{l,t})$ is satisfied under input module $\mathcal{A}_\mathrm{p}$. Accordingly, for any module $\mathcal{B}_\mathrm{p}$ and $\mathcal{C}_\mathrm{p}$, the output action satisfies
\begin{equation}
\setcounter{equation}{28}
(\mathcal{C}_\mathrm{p} \circ \mathcal{B}_\mathrm{p} \circ \mathcal{A}_\mathrm{p})(g\mathbf{o}_{l,t}) = (\mathcal{C}_\mathrm{p} \circ \mathcal{B}_\mathrm{p} \circ \mathcal{A}_\mathrm{p})(\mathbf{o}_{l,t}),
\end{equation}
which indicates that $\mathcal{C}_\mathrm{p}$ conforms to the definition of PI.
\end{IEEEproof}
\begin{coro}
If $\mathcal{D}_\mathrm{p}$ is not modified when $\mathcal{A}_\mathrm{p}$ becomes PI, it will immediately become PI, i.e.,
\begin{equation}
\setcounter{equation}{29}
(\mathcal{D}_\mathrm{p} \circ \mathcal{B}_\mathrm{p} \circ \mathcal{A}_\mathrm{p})(g\mathbf{o}_{l,t}) = (\mathcal{D}_\mathrm{p} \circ \mathcal{B}_\mathrm{p} \circ \mathcal{A}_\mathrm{p})(\mathbf{o}_{l,t}),
\end{equation}
and the output module $\mathcal{D}_\mathrm{p}$ complies with
\begin{equation}
\setcounter{equation}{30}
\mathcal{D}_\mathrm{p}(g\mathbf{o}_{l,t}) = \mathcal{D}_\mathrm{p}(\mathbf{o}_{l,t}).
\end{equation}
\end{coro}
\begin{thm}
To make $\mathcal{D}_\mathrm{p}$ become PE in the case of $\mathcal{A}_\mathrm{p}$ being PI, it is necessary to introduce $\mathbf{o}_{l,t}$ as an additional input, i.e., modifying the PE's output module in (26) as
\begin{equation}
\setcounter{equation}{31}
\pi_l(\mathbf{a}_{l,t,\mathrm{equiv}}|\mathbf{o}_{l,t}) = \mathcal{D}_\mathrm{p}(\mathbf{O}_{t},\mathbf{B}_{t})_l.
\end{equation}
\end{thm}
\begin{IEEEproof}
For the given $\mathcal{D}_\mathrm{p}$, which knows in advance the order of the given observation $\mathbf{o}_{l,t}$, then can derive $\mathcal{D}_\mathrm{p}$ as
\begin{equation}
\setcounter{equation}{32}
\begin{aligned}
\mathcal{D}_\mathrm{p}(g\mathbf{O}_{t},\mathbf{B}_{t})=g\pi_l(\mathbf{a}_{l,t,\mathrm{equiv}}|\mathbf{o}_{l,t}),
\end{aligned}
\end{equation}
which indicates that $\mathcal{D}_\mathrm{p}$ conforms to the definition of PE.
\end{IEEEproof}
\begin{rem}
Note that modifying four different modules $\{\mathcal{A}_\mathrm{p}, \mathcal{B}_\mathrm{p}, \mathcal{C}_\mathrm{p}, \mathcal{D}_\mathrm{p}\}$ in the permutation policy network appropriately can achieve
\begin{equation}
\setcounter{equation}{33}
\begin{aligned}
\pi_l(\mathbf{a}_{l,t}|g\mathbf{o}_{l,t}) &= g\pi_l(\mathbf{a}_{l,t}|\mathbf{o}_{l,t}), \quad \forall g \in \mathcal{G}, \forall \mathbf{o}_{l,t} \in \mathcal{O}_l \\
&=(\pi_l(\mathbf{a}_{l,t,\mathrm{inv}}|\mathbf{o}_{l,t}),g\pi_l(\mathbf{a}_{l,t,\mathrm{equiv}}|\mathbf{o}_{l,t})),
\end{aligned}
\end{equation}
with $\mathcal{B}_\mathrm{p}$ and $\mathcal{C}_\mathrm{p}$ remaining unchanged. Usually, we can incorporate recurrent neural network (RNN) or GNN into $\mathcal{B}_\mathrm{p}$ to handle the partially observable inputs. Therefore, following this minimum modification principle can ensure that implementing PI and PE without modifying the backbone architecture of the underlying MARL algorithm is beneficial.
\end{rem}
\subsection{Dynamic Permutation Network}
As a common practice adopted by many MARL algorithms, e.g., MADDPG and MAPPO, we can adopt a novel dynamic permutation network instead of the original actor networks to complete the action allocation for each agent, where $\mathcal{A}_\mathrm{p}$, $\mathcal{C}_\mathrm{p}$, and $\mathcal{D}_\mathrm{p}$ are the basic modules composed of fully connected (FC) layers and $\mathcal{B}_\mathrm{p}$ is a backbone module constructed by RNN or GNN architectures. Then, we denote
the set of weight matrices for the FC layer in $\mathcal{A}_\mathrm{p}$ as $\mathcal{W}_\mathrm{in}^\mathcal{A}$, then the output $\mathbf{A}_t$ of module $\mathcal{A}_\mathrm{p}$ in (26) can be represented as
\begin{equation}
\setcounter{equation}{34}
\mathbf{A}_t = \mathcal{A}_\mathrm{p}(\mathbf{O}_t) = \left[\mathbf{p}_{1,t},\ldots,\mathbf{p}_{L,t}\right]^T \in \mathbb{R}^{L \times d_\mathrm{r}}
\end{equation}
with the $l$-th element
\begin{equation}
\setcounter{equation}{35}
\mathbf{p}_{l,t} = \mathcal{A}_\mathrm{p}(\mathbf{O}_{l,t}^\mathrm{r}) = \sum_{l'=1}^{L}\left(\mathbf{W}_{l,t}^{\mathcal{A},l'}\right)^T\mathbf{o}_{l,t}^{l',\mathrm{r}} \in \mathbb{R}^{d_\mathrm{r} \times 1},
\end{equation}
where $\mathbf{O}_{l,t}^\mathrm{r}=[\mathbf{o}_{l,t}^{1,\mathrm{r}},\ldots,\mathbf{o}_{l,t}^{L,\mathrm{r}}]^T \in \mathbb{R}^{L \times f_\mathrm{r}}$ is the reshaping of observation $\mathbf{o}_{l,t} \in \mathbb{R}^{f \times 1}$ with $f = Lf^\mathrm{r}$, and $\mathbf{W}_{l,t}^{\mathcal{A},l'} \in \mathbb{R}^{f_\mathrm{r} \times d_\mathrm{r}}$ represent a weight matrix from $\mathcal{W}_\mathrm{in}^\mathcal{A}$ with the dimension of the hidden layer $d_\mathrm{r}$.
\begin{rem}
Note that to convert module $\mathcal{A}_\mathrm{p}$ into PI, a weight selection strategy needs to be introduced to assign weight matrices that are not affected by order to each $\mathbf{o}_{l,t}^{l',\mathrm{r}}$ in $\mathbf{o}_{l,t}$, so that the same $\mathbf{o}_{l,t}^{l',\mathrm{r}}$ will always be multiplied by the same weight matrix $\mathbf{W}_{l,t}^{\mathcal{A},l'}$. Then, the weight selection network with the one-hot encoding $p_\mathrm{in}(\hat{\mathbf{o}}_{l,t}^{l',\mathrm{r}})$ can be modeled based on the Gumbel-MAX estimator as
\begin{equation}
\setcounter{equation}{36}
p_\mathrm{in}\left((\mathbf{W}_{l,t}^{\mathcal{A},1},\ldots,\mathbf{W}_{l,t}^{\mathcal{A},L})\Big|\mathbf{o}_{l,t}^{l',\mathrm{r}}\right)=\mathcal{S}\left(\mathcal{M}(\mathbf{o}_{l,t}^{l',\mathrm{r}})\right),
\end{equation}
and the output of $\mathcal{A}_\mathrm{p}$ can be represented as
\begin{equation}
\setcounter{equation}{37}
\mathbf{p}_{l,t} = \mathcal{A}_\mathrm{p}(g\mathbf{O}_{l,t}^\mathrm{r}) = \sum_{l'=1}^{L}\left(p_\mathrm{in}(\hat{\mathbf{o}}_{l,t}^{l',\mathrm{r}})\mathcal{W}_\mathrm{in}^\mathcal{A}\right)^T(\hat{\mathbf{o}}_{l,t}^{l',\mathrm{r}}),
\end{equation}
where $p_\mathrm{in}(\mathbf{W}_{l,t}^{\mathcal{A},j}|\mathbf{o}_{l,t}^{l',\mathrm{r}})$ indicates the probability that $\mathbf{o}_{l,t}^{l',\mathrm{r}}$ selects the $j$-th weight matrix $\mathbf{W}_{l,t}^{\mathcal{A},j}$ from $\mathcal{W}_\mathrm{in}^\mathcal{A}$. Besides, $\mathcal{S}(\cdot)$ and $\mathcal{M}(\cdot)$ denote the softmax function and multi-layer perceptron, respectively.
Therefore, the resulting $\mathbf{A}_t$ remains the same regardless of the arranged orders of $\mathbf{o}_{l,t}^{l',\mathrm{r}}$ in $\mathbf{o}_{l,t}$, i.e., input module A becomes PI.
\end{rem}

Similarly, we denote $\mathbf{W}_{l,t}^{\mathcal{D}} \in \mathbb{R}^{d_\mathrm{r} \times d_\mathrm{a}}$ as a weight matrix from the set of weight matrices $\mathcal{W}_\mathrm{out}^\mathcal{D}$ with the dimension of the action $d_\mathrm{a}$, then the output $\pi_l(\mathbf{a}_{l,t,\mathrm{equiv}}|\mathbf{o}_{l,t})  \in \mathbb{R}^{d_\mathrm{a} \times 1}$ of module $\mathcal{D}_\mathrm{p}$ in (31) can be represented as
\begin{equation}
\setcounter{equation}{38}
\pi_l(\mathbf{a}_{l,t,\mathrm{equiv}}|\mathbf{o}_{l,t}) = \mathcal{D}_\mathrm{p}\left(\mathbf{O}_{t},\mathbf{B}_{t}\right)_l = \left(\mathbf{W}_{l,t}^{\mathcal{D}}\right)^T\mathbf{b}_{l,t},
\end{equation}
where $\mathbf{b}_{l,t} \in \mathbb{R}^{d_\mathrm{r} \times 1}$ denote the $l$-th element in the output $\mathbf{B}_{t}$.
\begin{rem}
Note that in order for module $\mathcal{D}_\mathrm{p}$ to implement PE, a weight selection network with the one-hot encoding $p_\mathrm{out}(g{\mathbf{O}}_{l,t}^{\mathrm{r}})$ still needs to be constructed for each entity-correlated action, such that the $j$-th element $\hat{\mathbf{o}}_{l,t}^{j,\mathrm{r}}$ in $\hat{\mathbf{O}}_{l,t}^{\mathrm{r}}$ will always correspond to the same vector of $\mathbf{w}_{l,t}^{j,\mathcal{D}}$ in $\mathbf{W}_{l,t}^{\mathcal{D}} \in \mathcal{W}_\mathrm{out}^\mathcal{D}$. Then, the $l$-th output $\pi_l(\mathbf{a}_{l,t,\mathrm{equiv}}|g\mathbf{O}_{l,t}^\mathrm{r})$ of $\mathcal{D}_\mathrm{p}$ will always have the same value, which can be represented as
\begin{equation}
\setcounter{equation}{39}
\begin{aligned}
\pi_l(\mathbf{a}_{l,t,\mathrm{equiv}}|g\mathbf{O}_{l,t}^\mathrm{r}) &= \mathcal{D}_\mathrm{p}\left(g\mathbf{O}_{t},\mathbf{B}_{t}\right)_l\\
&=\left(p_\mathrm{out}(g{\mathbf{O}}_{l,t}^{\mathrm{r}})\mathcal{W}_\mathrm{out}^\mathcal{D}\right)^T\mathbf{b}_{l,t}.
\end{aligned}
\end{equation}
Therefore, the input order of $\hat{\mathbf{o}}_{l,t}^{j,\mathrm{r}}$ in $\hat{\mathbf{O}}_{l,t}^{\mathrm{r}}$ change will result in the same output order change, thus achieving PE.
\end{rem}
\begin{algorithm}[t]
\label{algo:AIRMN}
\caption{Dynamic Hyper Permutation Network }
    \KwIn{Observation information $\mathbf{O}_{t}^{(i)} = [\mathbf{o}_{1,t},\ldots,\mathbf{o}_{L,t}]^T$;}
    \KwOut{Optimal action $\mathbf{a}_{l,t}^{(i)}$, including $\mathbf{a}_{l,t,\mathrm{inv}}^{(i)}$ for PI's actions and $\mathbf{a}_{l,t,\mathrm{equiv}}^{(i)}$ for PE's actions;}

    {\bf Initiation:} Initial number $i=0$; Hidden state matrix $\mathbf{R}_{t}^{(i)} = \mathbf{0}_{L \times d_r}$; Maximum iteration number $I_{\max}$;\\

    \Repeat(){$r_{\mathrm{ex},t}^{(i)}<r_{\mathrm{ex},t}^{(i)}$ or $i \geqslant I_{\max}$}
        {
        $i=i+1$\\
        Module $\mathcal{A}_\mathrm{p}$: Update the output state matrix $\mathbf{A}_t^{(i)}$ of the initial PI architecture with $\mathbf{O}_{t}^{(i)}$ based on (40);\\
        Module $\mathcal{B}_\mathrm{p}$: Update the output state matrix $\mathbf{B}_t^{(i)}$ of RNN architecture with $\mathcal{R}_{\mathrm{a}}(\mathbf{A}_t^{(i)},\mathbf{B}_t^{(i-1)})$, and update the next hidden state $\mathbf{R}_t^{(i)}=\mathbf{B}_t^{(i)}$;\\
        Module $\mathcal{C}_\mathrm{p}$: Obtain PI's actions $\mathbf{a}_{l,t,\mathrm{inv}}^{(i)}$ with $\mathbf{B}_t^{(i)}$ under invariant architecture based on (26);\\
        Module $\mathcal{D}_\mathrm{p}$: Obtain PE's actions $\mathbf{a}_{l,t,\mathrm{equiv}}^{(i)}$ with $\mathbf{O}_{t}^{(i)}$ and $\mathbf{B}_t^{(i)}$ under equivariant architecture based on (41);\\
        Update sum reward $r_{\mathrm{ex},t}^{(i)} = \sum_{k=1}^{K}\mathrm{SE}_{k,t}^{(i)}$ with generated actions $[\mathbf{a}_{l,t,\mathrm{inv}}^{(i)},\mathbf{a}_{l,t,\mathrm{equiv}}^{(i)}]$;
        }
\end{algorithm}
\begin{figure*}[t]
\centering
    \includegraphics[scale=0.26]{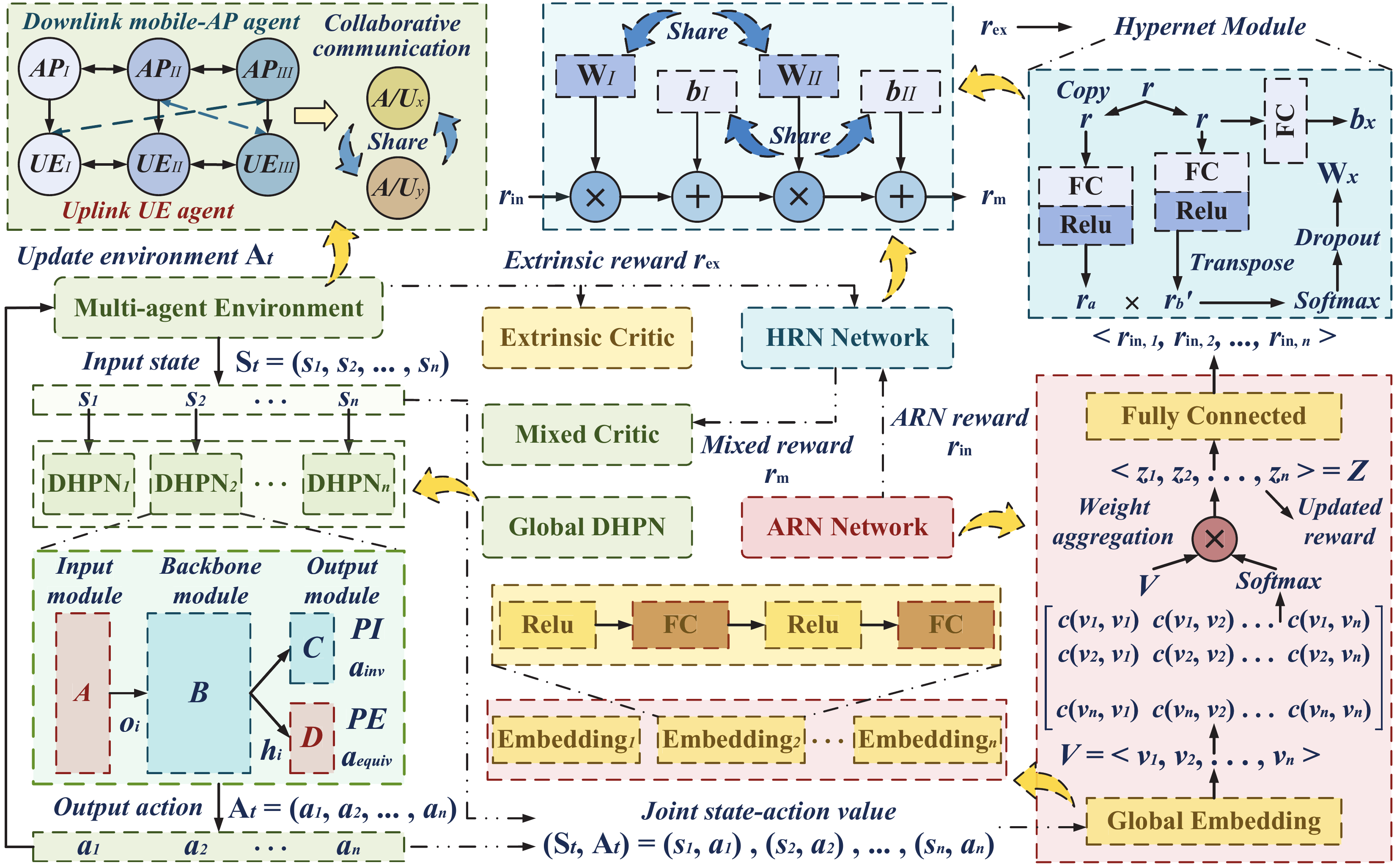}
    \caption{The overview of a scalable framework for joint mobility and downlink power control under cell-free mMIMO systems, including a GNN-aided communication architecture, a permutation architecture, and a directional decoupling architecture.
    \label{fig1}}
\end{figure*}
\subsection{Dynamic Hyper Permutation Network}
Considering that in a dynamic permutation network, we can only select weight matrices for each $\hat{\mathbf{o}}_{l,t}^{l',\mathrm{r}}$ from these finite sets $\mathcal{W}_\mathrm{in}^\mathcal{A}$ and $\mathcal{W}_\mathrm{out}^\mathcal{D}$, and their limited size may result in weight matrices assigned to each $\hat{\mathbf{o}}_{l,t}^{l',\mathrm{r}}$ not being the best fit. In contrast, a hypernetwork can provide an infinite number of candidate weight matrices and generate customized embedding weights for each $\hat{\mathbf{o}}_{l,t}^{l',\mathrm{r}}$, so that the solution space of the assignment is no longer constrained.

Correspondingly, we construct two different hypernetworks $\mathcal{M}_{\mathrm{in}}(\mathcal{W}_{\mathrm{in},\mathrm{h}}^\mathcal{A}|\hat{\mathbf{o}}_{l,t}^{l',\mathrm{r}})$ and $\mathcal{M}_{\mathrm{out}}(\mathcal{W}_{\mathrm{out},\mathrm{h}}^\mathcal{D}|g{\mathbf{O}}_{l,t}^{\mathrm{r}})$ for modules $\mathcal{A}_\mathrm{p}$ and $\mathcal{D}_\mathrm{p}$, respectively, to generate corresponding weight matrices for each $\hat{\mathbf{o}}_{l,t}^{l',\mathrm{r}}$. Then, the output $\mathbf{p}_{l,t}$ of $\mathcal{A}_\mathrm{p}$ can be modeled with the matrix generated by the hypernetwork $\mathcal{M}_{\mathrm{in}}(\mathcal{W}_{\mathrm{in},\mathrm{h}}^\mathcal{A}|\hat{\mathbf{o}}_{l,t}^{l',\mathrm{r}})$ as
\begin{equation}
\setcounter{equation}{40}
\mathbf{p}_{l,t} = \mathcal{A}_\mathrm{p}(g\mathbf{O}_{l,t}^\mathrm{r}) = \sum_{l'=1}^{L}\left(\mathcal{M}_{\mathrm{in}}(\mathcal{W}_{\mathrm{in},\mathrm{h}}^\mathcal{A}|\hat{\mathbf{o}}_{l,t}^{l',\mathrm{r}})\right)^T(\hat{\mathbf{o}}_{l,t}^{l',\mathrm{r}}),
\end{equation}
and the output of $\mathcal{D}_\mathrm{p}$ can be modeled with the matrix generated by the hypernetwork $\mathcal{M}_{\mathrm{out}}(\mathcal{W}_{\mathrm{out},\mathrm{h}}^\mathcal{D}|g{\mathbf{O}}_{l,t}^{\mathrm{r}})$ as
\begin{equation}
\setcounter{equation}{41}
\begin{aligned}
\pi_l(\mathbf{a}_{l,t,\mathrm{equiv}}|g\mathbf{O}_{l,t}^\mathrm{r}) &= \mathcal{D}_\mathrm{p}\left(g\mathbf{O}_{t},\mathbf{B}_{t}\right)_l\\
&=\left(\mathcal{M}_{\mathrm{out}}(\mathcal{W}_{\mathrm{out},\mathrm{h}}^\mathcal{D}|g{\mathbf{O}}_{l,t}^{\mathrm{r}})\right)^T\mathbf{b}_{l,t}.
\end{aligned}
\end{equation}
\begin{rem}
Note that different $\hat{\mathbf{o}}_{l,t}^{l',\mathrm{r}}$ can generate different weight matrices and the same $\hat{\mathbf{o}}_{l,t}^{l',\mathrm{r}}$ will always correspond to the same one no matter where it is arranged. Moreover, since each $\hat{\mathbf{o}}_{l,t}^{l',\mathrm{r}}$ is separately embedded by its corresponding weight matrix $\mathcal{M}_{\mathrm{in}}(\mathcal{W}_{\mathrm{in},\mathrm{h}}^\mathcal{A}|\hat{\mathbf{o}}_{l,t}^{l',\mathrm{r}})$ and merged through the ``element-wise sum" function, the PI property is ensured.
\end{rem}

Moreover, the overall process of the dynamic hyper permutation network is shown in \textbf{Algorithm 1}.
\section{Scalable Framework}
In this section, we first introduce GNNs and illustrate how to apply them to solve collaborative optimization problems between agents. Then, we develop a scalable framework with MARL architecture to address the challenges faced by conventional optimization schemes in cell-free mMIMO systems.
\subsection{GNN-aided Communication Architecture}
Considering the high system overhead and computational complexity of global communication in cell-free mMIMO systems, unreasonable local communication is not to improve system performance. By contrast, introducing a GNN-aided communication architecture to select suitable neighboring agents for collaborative communication and effectively aggregate received messages is a promising solution \cite{[20],[21]}, as shown in Module $\mathcal{B}_\mathrm{p}$ in Fig. 2. Then, we can model the cell-free mMIMO system as a graph with an ordered pair $\mathcal{G}_{\mathrm{c}}=<\mathcal{V},\mathcal{E}>$, where $\mathcal{V}$ represents the set of nodes, e.g., mobile-APs or UEs, and $\mathcal{E} \subset {\{(v,u)|v,u \in \mathcal{V}\}}$ denotes the set of edges associated with node pairs $(v,u)$. In general, GNNs extend the spatial convolution from convolutional neural networks to graphs, where each node $l$ updates its hidden state $\mathbf{s}_{\mathrm{h},l,t}^{(j)}$ in the $j$-th layer based on aggregated information from its neighboring nodes.
Specifically, the update process of each GNN layer consists of two stages, as follows
\begin{itemize}
\item \textbf{Aggregation Network}: The first step is to adopt an aggregator $\mathcal{A}_\mathrm{h}(\cdot)$ for each node $l$ to collect the hidden states $\mathbf{s}_{\mathrm{h},l',t}^{(j-1)}$ of nodes $l'$ from neighboring set $\mathcal{N}_{l,t}$ at $j$-th layer, i.e., $l' \in \mathcal{N}_{l,t}$, which can be represented as
\begin{equation}
\setcounter{equation}{42}
\mathbf{a}_{\mathrm{h},l,t}^{(j)}= \mathcal{L}\Big(\mathcal{A}_\mathrm{h}\Big(\mathcal{E}_\mathrm{o}(\mathbf{o}_{l,t}),\mathbf{s}_{\mathrm{h},l,t}^{(j-1)},\mathbf{s}_{\mathrm{h},l',t}^{(j-1)}\Big|\mathbf{W}_{\mathrm{a},l,t}^{(j)}\Big)\Big),
\end{equation}
where $\mathcal{L}(\cdot)$ and $\mathcal{E}_\mathrm{o}(\cdot)$ denote a pooling function (e.g., max or sum pooling) and a state encoder, respectively. Besides, $\mathbf{W}_{\mathrm{a},l,t}^{(j)}$ represents the learnable weight in the aggregated network to selectively aggregate hidden states within the neighborhood.

\item \textbf{Combination Network}: The second step is to adopt a combiner $\mathcal{C}_\mathrm{h}(\cdot)$ for each node $l$ to effectively handle the aggregated state $\mathbf{a}_{\mathrm{h},l,t}^{(j)}$ and its own state at $j$-th layer, then the updated hidden state $\mathbf{s}_{\mathrm{h},l,t}^{(j)}$ can be represented as
\begin{equation}
\setcounter{equation}{43}
\mathbf{s}_{\mathrm{h},l,t}^{(j)} = \mathcal{C}_\mathrm{h}\Big(\mathcal{E}_\mathrm{o}(\mathbf{o}_{l,t}),\mathbf{s}_{\mathrm{h},l,t}^{(j-1)},\mathbf{a}_{\mathrm{h},l,t}^{(j)}\Big|\mathbf{W}_{\mathrm{c},l,t}^{(j)}\Big),
\end{equation}
where $\mathbf{W}_{\mathrm{c},l,t}^{(j)}$ is the learnable weight in the combination network to selectively combine aggregated and self states.
\end{itemize}

After iteratively updating all hidden states through $J$ GNN layers, we can adopt a message encoder $\mathcal{E}_\mathrm{m}(\cdot)$ to encode all updated hidden states as received messages, which can be represented as
\begin{equation}
\setcounter{equation}{44}
(\mathbf{m}_{\mathrm{h},l,t},\mathbf{G}_{\mathrm{h},t}) = \mathcal{E}_\mathrm{m}\Big(\mathbf{s}_{\mathrm{h},l,t}^{(J)},\mathbf{G}_{\mathrm{h},t-1}\Big),
\end{equation}
where $\mathbf{m}_{\mathrm{h},l,t}$ denotes the information received by each agent $l$, and $\mathbf{G}_{\mathrm{h},t}$ represents the adjacency matrix between nodes, which can update the neighboring set $\{\mathcal{N}_{1,t},\ldots,\mathcal{N}_{L,t}\}$ at time slot $t$ to $\{\mathcal{N}_{1,t+1},\ldots,\mathcal{N}_{L,t+1}\}$ at time slot $t + 1$.

Correspondingly, we can combine the received messages $\mathbf{m}_{\mathrm{h},l,t}$ to update the self observation $\mathbf{o}_{l,t}$ of each agent $l$, which can be modeled as
\begin{equation}
\setcounter{equation}{45}
\mathbf{o}_{l,t} = \mathcal{T}_{\mathrm{m}}(\mathbf{o}_{l,t},\mathbf{m}_{\mathrm{h},l,t}),
\end{equation}
where $\mathcal{T}_{\mathrm{m}}{(\cdot)}$ denotes the state update function.
\subsection{MARL-based Decoupling Architecture}
Aligning with the development trend of joint mobility and downlink power control in cell-free mMIMO systems, as shown in Table \uppercase\expandafter{\romannumeral1},
we propose a MARL-based scalable framework, namely SF-MADDPG. This framework connects GNN-aided communication architecture, permutation architecture, and directional decoupling architecture, as shown in Fig. 3.

In the proposed framework, considering that the joint problem in (22) belongs to the downlink, we can define mobile-AP and UE as homogeneous downlink agents and heterogeneous uplink agents, respectively, where the number of studied agents $L$ is equal to the number of mobile-APs $M$, satisfying the index $l = m$.
It is worth noting that to mitigate the impact of outdated decisions in the MARL environment, we deploy more antennas at each mobile-AP to converge the normalized instantaneous channel gain to a determined average channel gain. On the other hand, we introduce a slower learning rate in the original network to ensure that transmission power and mobility are always optimized for instantaneous system conditions.
Moreover, the observation $\mathbf{o}_{l,t}$ of each downlink agent $l$ at time slot $t$ is the combination of collective channel state information $\mathbf{H}_l=[\mathbf{h}_{l,1},\ldots,\mathbf{h}_{l,K}]^T$, observed large-scale fading coefficients $\boldsymbol{\beta}_l=[\beta_{l,1},\ldots,\beta_{l,K}]^T$, and perceived coordinate information $\mathbf{C}_l$, while the assigned actions $\mathbf{a}_{l,t}$ are the combination of mobility actions $\mathbf{a}_{\mathrm{m},l,t}$, overall AP power actions $\mathbf{a}_{\mathrm{sp},l,t}$, and antenna power actions $\mathbf{a}_{\mathrm{mp},l,t}$.
Then, according to the permutation principles of (23) and (24), we can divide mobility actions $\mathbf{a}_{\mathrm{m},l,t}$ into PI properties, while agent power actions $\mathbf{a}_{\mathrm{sp},l,t}$ and antenna power actions $\mathbf{a}_{\mathrm{mp},l,t}$ into PE properties, as shown in Fig. 4, which satisfies
\begin{equation}
\setcounter{equation}{46}
\begin{split}
\left \{
\begin{array}{ll}
    \text{PI's output: } \pi_l(\mathbf{a}_{\mathrm{m},l,t}|\mathbf{o}_{l,t}) = \mathcal{C}_\mathrm{p}(\mathbf{B}_{t})_l,\\
    \text{PE's output: } \pi_l(\mathbf{a}_{\mathrm{sp},l,t}|\mathbf{o}_{l,t}) = \mathcal{D}_\mathrm{p}^\mathrm{sp}(g\mathbf{O}_{t},\mathbf{B}_{t})_l,\\
    \text{PE's output: } \pi_l(\mathbf{a}_{\mathrm{mp},l,t}|\mathbf{o}_{l,t}) = \mathcal{D}_\mathrm{p}^\mathrm{mp}(g\mathbf{O}_{t},\mathbf{B}_{t})_l.
\end{array}
\right.
\label{eq45}
\end{split}
\end{equation}
\begin{figure}[t]
\centering
    \includegraphics[scale=0.2125]{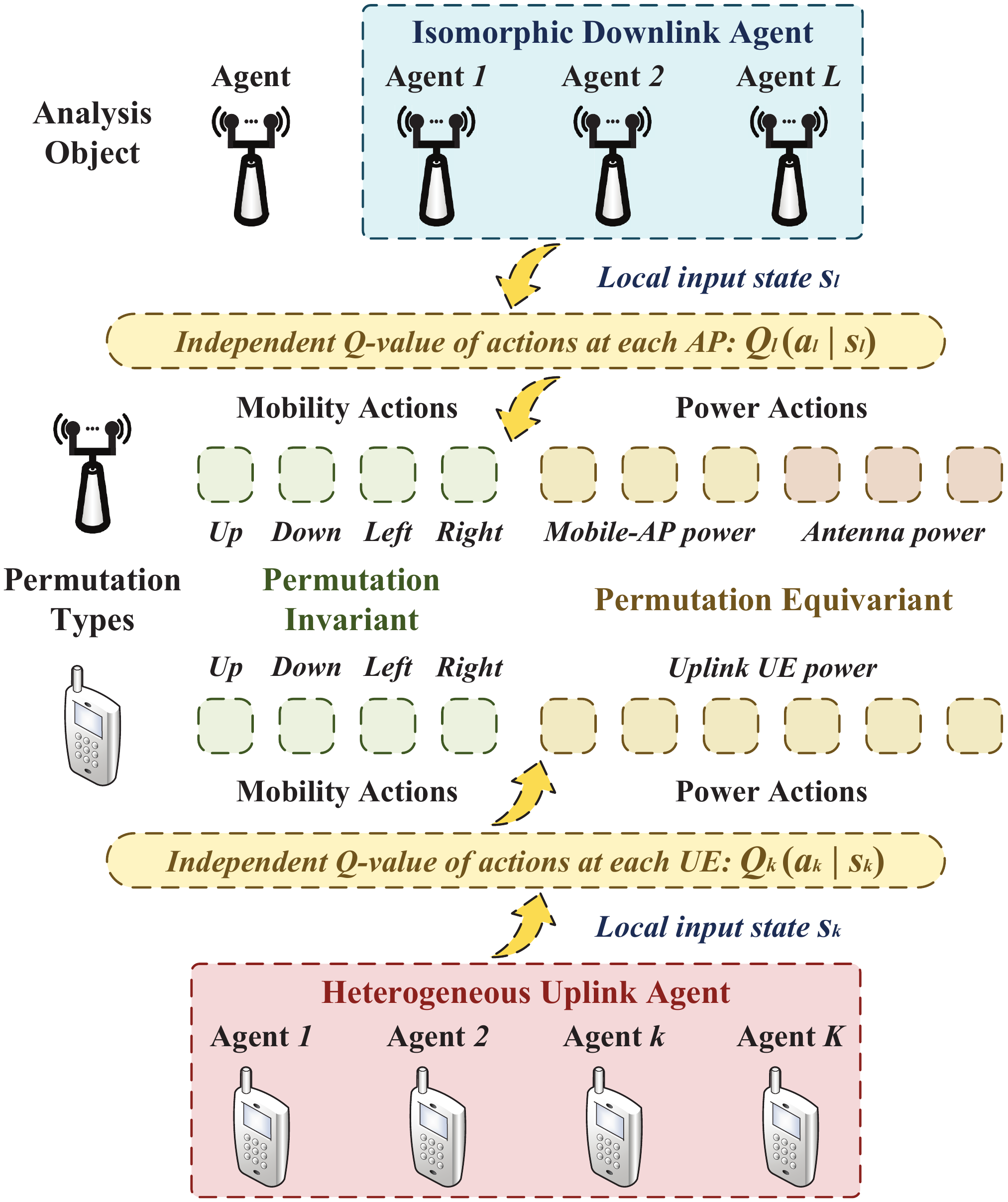}
    \caption{The two permutation types of actions in the cell-free mMIMO system include permutation equivariant, such as mobility actions and agent power actions, as well as permutation invariant, such as antenna power actions.
    \label{fig1}}
\end{figure}
\begin{algorithm}[t]
\label{algo:AIRMN}
\caption{MARL-based Scalable Framework}
    \KwIn{Collective channel state information $\mathbf{H}_l$ and large-scale fading information $\boldsymbol{\beta}_l$; Perceived coordinate information $\mathbf{C}_{l}$;}
    \KwOut{Optimal extrinsic reward $r_{\mathrm{ex},t} = \sum_{k=1}^K\mathrm{SE}_{k,t}$;}
    {\bf Initiation:} Initial time slot $t=0$; Number of episodes $E_{\max}$; Maximum tolerable fading interval $t_{\max}$; Empty network experience extraction pool $\mathcal{P}$;\\
    \Repeat(){$t \geqslant E_{\max}$ or \{$r_{\mathrm{ex},c}<r_{\mathrm{ex},c+1}$, $t - t_{\max} \leqslant c \leqslant t$\}}
        {
        $t=t+1$\\
        Generate collective state $\mathbf{s}_{l,t}$ and observation $\mathbf{o}_{l,t}$ for all APs with observed information $\mathbf{H}_l$, $\boldsymbol{\beta}_l$, and $\mathbf{C}_{l}$;\\
        Obtain action information $\mathbf{a}_{l,t}$ generated by HDPN architecture with $\mathbf{o}_{l,t}$ based on \textbf{Algorithm 1};\\
        Generate extrinsic reward $\mathbf{r}_{\mathrm{ex},l,t}$, next state $\mathbf{s}_{l,t+1}$, and observation $\mathbf{o}_{l,t+1}$ after all agents interact with the environment using the obtained actions $\mathbf{a}_{l,t}$;\\
        Update intrinsic reward $\mathbf{r}_{\mathrm{in},t} = \mathcal{A}_{r}(\mathbf{O}_{t},\mathbf{A}_{t})$ and mixed reward $\mathbf{r}_{\mathrm{m},t} = \mathcal{H}_{r}(\mathbf{r}_{\mathrm{in},t},\mathbf{r}_{\mathrm{ex},t})$;\\
        Store $<\mathbf{o}_{l,t}, \mathbf{s}_{l,t}, \mathbf{a}_{l,t}, \mathbf{r}_{\mathrm{ex},t}, \mathbf{r}_{\mathrm{m},t}, \mathbf{s}_{l,t+1}>$ to $\mathcal{P}$;\\
        \If(){update the MARL architecture}
            {
            Sample a mini-batch $\mathcal{P}_s$ from $\mathcal{P}$ randomly;\\
            Calculate the mixed critic value with individual experience $\mathcal{P}_{s,l}$ for all agents based on (48);\\
            Update all mixed critic networks;\\
            Calculate the policy gradient of all HDPN based on (49) and update them based on (51);\\
            Calculate the extrinsic critic value with collective experience $\mathcal{P}_s$ based on (52) and update it;\\
            Calculate the policy gradient of ARN and HRN architectures based on (53) and update them;\\
            }
        }
\end{algorithm}

Moreover, considering that the reward value in the downlink joint optimization problem is summarized at all mobile-APs, while the SE value in the cell-free mMIMO is summarized at all UEs, such misalignment makes it impossible to allocate the obtained SE value to each mobile-AP according to a reasonable contribution. By contrast, using credit assignment can effectively address this challenge of misalignment \cite{[25],[26],[27]}. Specifically, an attention-based intrinsic reward network (ARN) $\mathcal{A}_\mathrm{r}(\cdot)$ is set for each agent, which provides different incentives $\mathbf{r}_{\mathrm{in},t}=[r_{\mathrm{in},1,t},\ldots,r_{\mathrm{in},L,t}]^T$ to the agent at each time slot $t$ to guide the updating of their strategies, and is combined with extrinsic rewards $\mathbf{r}_{\mathrm{ex},t}=[r_{\mathrm{ex},1,t},\ldots,r_{\mathrm{ex},L,t}]^T$ to better divide rewards into contributions for each agent. Then, the updated reward $\mathbf{r}_{\mathrm{m},t}=[r_{\mathrm{m},1,t},\ldots,r_{\mathrm{m},L,t}]^T$ can be represented under the hypernetwork-based mixed reward network (HRN) $\mathcal{H}_\mathrm{r}(\cdot)$ as
\begin{equation}
\setcounter{equation}{47}
\mathbf{r}_{\mathrm{m},t}=\mathcal{H}_\mathrm{r}(\mathbf{r}_{\mathrm{in},t},\mathbf{r}_{\mathrm{ex},t})=\mathcal{H}_\mathrm{r}(\mathcal{A}_\mathrm{r}(\mathbf{O}_t,\mathbf{A}_t),\mathbf{r}_{\mathrm{ex},t}).
\end{equation}
The design of networks $\mathcal{A}_\mathrm{r}(\cdot)$ and $\mathcal{H}_\mathrm{r}(\cdot)$ is shown in Fig. 3.

\begin{figure*}[t]
\centering
    \includegraphics[scale=0.4125]{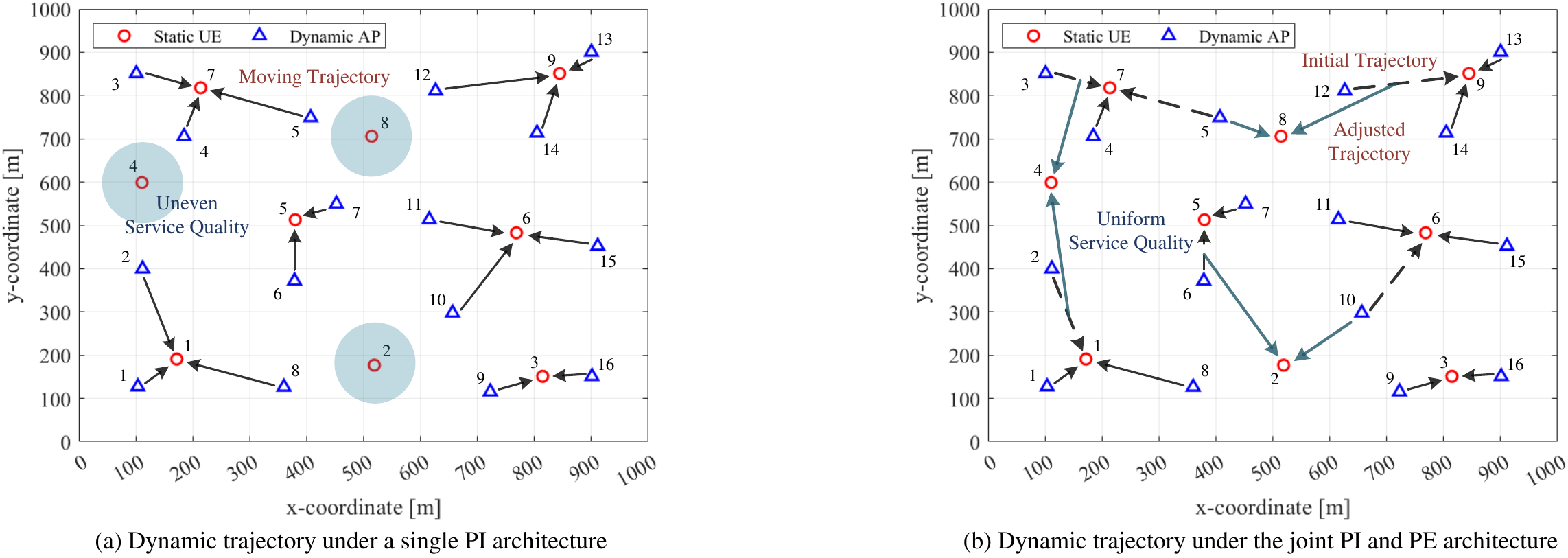}
    \vspace{-0.1cm}
    \caption{Dynamic trajectories under two different architectures with $M=16$, $K=9$, $N=8$, $\tau_p=K$, and $\Delta = \lambda/2$, including the single PI architecture in Fig. 5 (a), where the black solid arrow represents the movement trajectory of each mobile-AP, and the joint PI and PE architecture in Fig. 5 (b), where the black dashed arrow represents the initial movement trajectory of each mobile-AP, and the blue solid arrow represents the adjusted movement trajectory.
    \label{fig1}}
\end{figure*}
Besides, based on the mean-squared Bellman error function, the loss function of a mixed critic network can be modeled as
\begin{equation}
\setcounter{equation}{48}
L_{\mathrm{m}}(\theta_{Q_l}) = \mathbb{E}_{\mathbf{o}_{l,t},\mathbf{a}_{l,t}\sim \mathcal{P}}
\left[\left(Q_l(\mathbf{o}_{l,t},\mathbf{a}_{l,t})-y_{\mathrm{m},t,l}\right)^2\right],
\end{equation}
where $y_{\mathrm{m},t,l} = r_{\mathrm{m},t,l} + \gamma_{\mathrm{m}}Q_l(\mathbf{o}_{l,t}',\mathbf{a}_{l,t}')$ is the target value with the reward discount factor $\gamma_{\mathrm{m}}$ under the mixed critic network. Then, the policy gradient of each dynamic hyper permutation network (DHPN) $\theta_l$ can be defined as
\begin{equation}
\setcounter{equation}{49}
\nabla_{\theta_l}J_{\mathrm{m}}(\theta_l)=
{\mathbb{E}\left[\nabla_{\theta_l}\log\pi_{\theta_l}(\mathbf{a}_{l,t}|\mathbf{o}_{l,t})A_{\pi_{\theta_l}}(\mathbf{o}_{l,t},\mathbf{a}_{l,t})\right]}
\end{equation}
with the advantage function
\begin{equation}
\setcounter{equation}{50}
\begin{aligned}
A_{\pi_{\theta_l}}(\mathbf{o}_{l,t},\mathbf{a}_{l,t})&=\mathcal{R}(\mathbf{s}_{l,t},\mathbf{a}_{l,t}) + \gamma_{\mathrm{a}} V_{\pi_{\theta_l}}(\mathbf{o}_{l,t})-V_{\pi_{\theta_l}}(\mathbf{o}_{l,t}')\\
&= r_{\mathrm{ex},l,t} + \gamma_{\mathrm{a}} V_{\pi_{\theta_l}}(\mathbf{o}_{l,t})-V_{\pi_{\theta_l}}(\mathbf{o}_{l,t}'),
\end{aligned}
\end{equation}
where $V_{\pi_{\theta_l}}(\cdot)$ denotes the value function, and $\gamma_{\mathrm{a}}$ is the discount factor under the advantage function. Based on this, the policy  $\theta_l$ with a soft update rate $\tau_{\mathrm{a}}$ can be updated to
\begin{equation}
\setcounter{equation}{51}
\theta_l' \leftarrow \theta_l + \tau_{\mathrm{a}}\nabla_{\theta_l}\log\pi_{\theta_l}(\mathbf{a}_{l,t}|\mathbf{o}_{l,t})A_{\pi_{\theta_l}}(\mathbf{o}_{l,t},\mathbf{a}_{l,t}).
\end{equation}

Similarly, the loss function $L_{\mathrm{ex}}(\theta_{Q_\mathrm{g}})$ of an extrinsic critic network can be modeled based on collective observations $\mathbf{O}_t$ and actions $\mathbf{A}_t$ as
\begin{equation}
\setcounter{equation}{52}
L_{\mathrm{ex}}(\theta_{Q_\mathrm{g}}) = \mathbb{E}_{\mathbf{O}_{t},\mathbf{A}_{t}\sim \mathcal{P}}
\left[\left(Q_\mathrm{g}(\mathbf{O}_{t},\mathbf{A}_{t})-y_{\mathrm{ex},t}\right)^2\right],
\end{equation}
where $y_{\mathrm{ex},t} = r_{\mathrm{ex},t} + \gamma_{\mathrm{ex}}Q_\mathrm{g}(\mathbf{O}_{t}',\mathbf{A}_{t}')$ is the target value with the total extrinsic reward $r_{\mathrm{ex},t}=\sum_{l=1}^{L}r_{\mathrm{ex},l,t}=\sum_{k=1}^K\text{SE}_{k,t}$ and the discount factor $\gamma_{\mathrm{ex}}$. Moreover, the policy $\eta_l$ of ARN architecture can be updated by
\begin{equation}
\setcounter{equation}{53}
\begin{aligned}
\nabla_{\eta_l}\theta_l'&= \nabla_{\eta_l}\left[\theta_l + \tau_{\mathrm{a}}\nabla_{\theta_l}\log\pi_{\theta_l}(\mathbf{a}_{l,t}|\mathbf{o}_{l,t})A_{\pi_{\theta_l}}(\mathbf{o}_{l,t},\mathbf{a}_{l,t})\right]\\
&=\tau_{\mathrm{a}}\nabla_{\theta_l}\log\pi_{\theta_l}(\mathbf{a}_{l,t}|\mathbf{o}_{l,t})\nabla_{\eta_l}A_{\pi_{\theta_l}}(\mathbf{o}_{l,t},\mathbf{a}_{l,t}).
\end{aligned}
\end{equation}

Correspondingly, the overall process of the proposed framework is shown in \textbf{Algorithm 2}.

\begin{table}[t]
\centering
\fontsize{9}{9}\selectfont
\caption{The Model Parameter Settings in Our Experiments.}
\label{paper}
\begin{tabular}{ccc}
\toprule
\bf Parameters &  \bf Size \\
\midrule
1st hidden layer (MADDPG) & 128, Leaky Relu (0.01)\\
2nd hidden layer (MADDPG) & 64, Leaky Relu (0.01) \\
Hidden layer (DHPN and GNN) & 128 and 64 Relu \\
Hidden layer (RNN) & 256, Relu \\
Number of heads (DHPN and GNN) & 8 and 4  \\
Discounted factor $\gamma$, $\gamma_{\mathrm{a}}$, and $\gamma_{\mathrm{ex}}$ & 0.99 \\
Experience extraction pool $\mathcal{P}_s$ & 1024 \\
Maximum gradient clipping value $\xi$ & 0.5 \\
Soft update rate and learning rate & 0.01\\
\bottomrule
\end{tabular}
\end{table}
\section{Numerical Results}
In this section, simulation results are presented to evaluate the performance of the proposed SF-MADDPG algorithm in the mobile cell-free system, where all mobile-APs and UEs are assumed to be randomly distributed in a square area of $\text{1} \times \text{1}$ $\text{km}^2$ with a wrap-around scheme \cite{[5]}.  The pathloss model is computed by the COST 321 Walﬁsh-Ikegami model. Then, we consider communication with 20 MHz bandwidth, $\sigma^2 =$ -94 dBm noise power, and maximum downlink transmit power with $P_{\mathrm{ap},\max} = $ 1 W per mobile-AP and $P_{\mathrm{an},\max} = P_{\mathrm{ap},\max} / N$ per antenna. Each coherence block contains $\tau_c$ = 200 channel uses and $\tau_p = K$. Besides, the experimental details of the proposed SF-MADDPG algorithm are shown in Table \uppercase\expandafter{\romannumeral2}, and the simulation works are conducted with an Nvidia GeForce GTX 3060 Graphics Processing Unit.

Next, the performance of \textbf{Algorithm 2} is compared with the following benchmark schemes: (1) Heuristic-based schemes: Fractional \cite{[16]}. (2) MARL-based schemes: Permutation-based MADDPG (Pe-MADDPG) \cite{[23]}, Collaborative communication-based MADDPG (Co-MADDPG) \cite{[20]}, Attention-based Intrinsic Reward Mixing Network (AIRMN) \cite{[27]}, Learning Individual Intrinsic Reward Network (LIIR) \cite{[25]}, and basic MADDPG \cite{[16]}. (3) Multidimensional graph neural network (MDGNN)-aided schemes: one-dimensional (1D) MDGNN, two-dimensional (2D) MDGNN, and three-dimensional (3D) MDGNN \cite{[36]}, which can achieve superior system performance similar to conventional centralized schemes with reduced computational complexity. Specifically, for MDGNN (1D) schemes, permutation involves a single coordinate index of the AP, UE, or antenna, while MDGNN (2D) and MDGNN (3D) schemes effectively permute the joint order of the two or three mentioned above, thereby achieving better performance by fully utilizing permutation properties, but also increasing the required computational complexity.

Moreover, given the dynamically evolving nature of the wireless environment, it is imperative to continuously monitor the proposed SF-MADDPG algorithm in real-time throughout the training process. Therefore, we can compare the performance of the three baseline algorithms mentioned above in simulations to determine whether the trained model is outdated. Correspondingly, due to the serious impact of outdated models on system performance, we can also introduce an online learning mechanism in the proposed framework. This empowers all agents to adapt their policies based on current wireless conditions, ensuring that the trained model remains up-to-date in an ever-changing wireless environment and achieves superior system performance.
\begin{figure}[t]
\centering
    \includegraphics[scale=0.5]{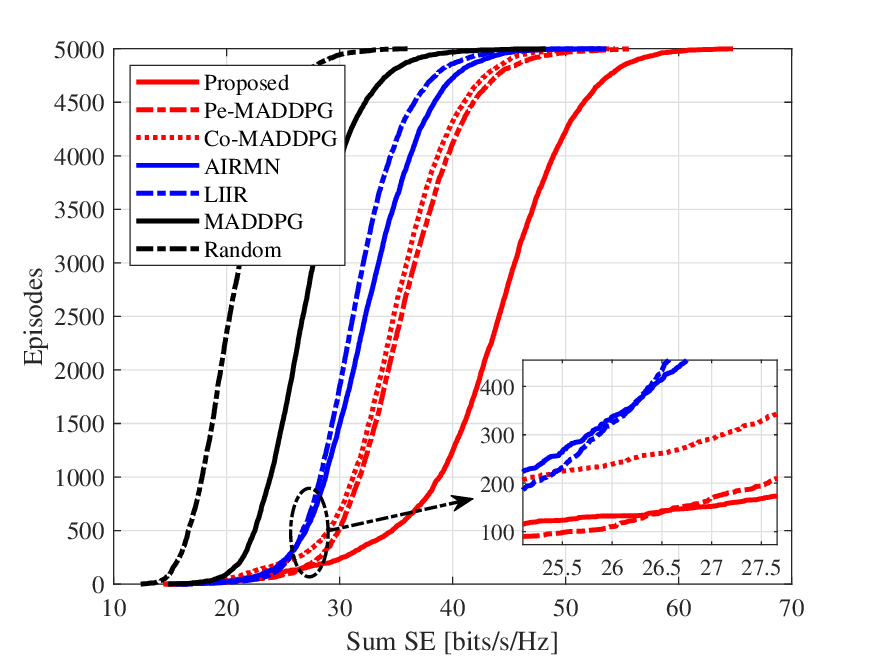}
    \vspace{-0.1cm}
    \caption{CDF of sum SE under a single PI architecture over different optimization schemes with $M=9$, $K=6$, $N=8$, $\tau_p=K$, and $\Delta = \lambda/2$.
    \label{fig1}}
\end{figure}
\begin{figure}[t]
\centering
    \vspace{-0.3cm}
    \includegraphics[scale=0.5]{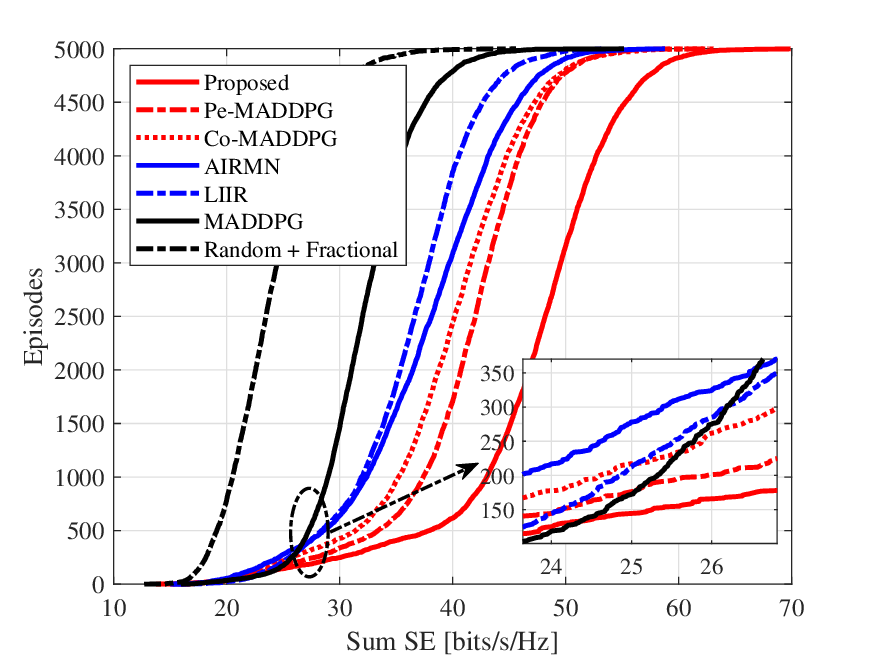}
    \vspace{-0.1cm}
    \caption{CDF of sum SE under the joint PI and PE architecture over different optimization schemes with $M=9$, $K=6$, $N=8$, $\tau_p=K$, and $\Delta = \lambda/2$.
    \label{fig1}}
\end{figure}
\begin{figure}[t]
\centering
    \includegraphics[scale=0.5]{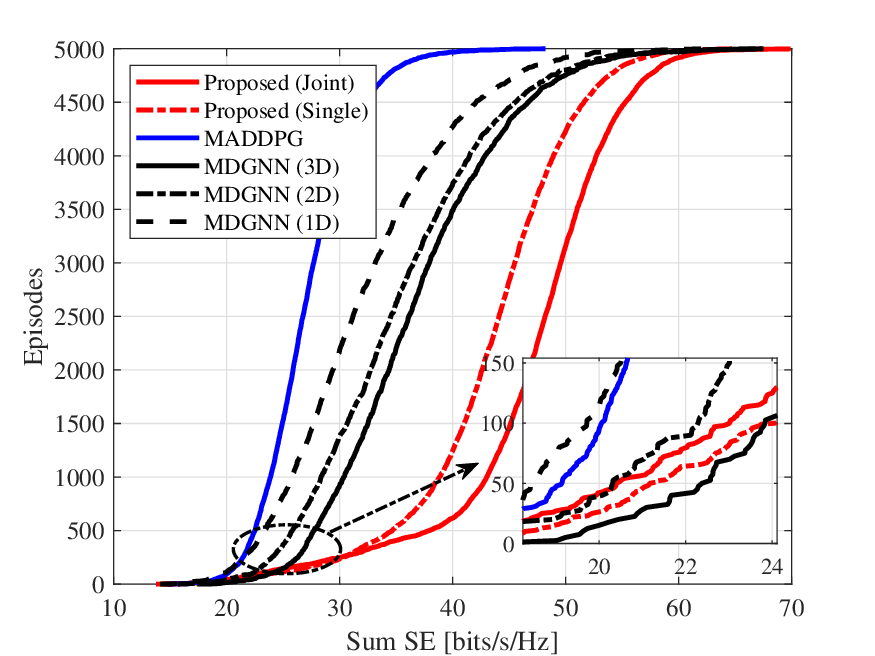}
    \vspace{-0.1cm}
    \caption{CDF of sum SE over different optimization schemes and architectures with $M=9$, $K=6$, $N=8$, $\tau_p=K$, and $\Delta = \lambda/2$.
    \label{fig1}}
\end{figure}

Fig. 5 investigates the moving trajectories of all agents under two different architectures, including a single PI architecture that only considers mobility actions and a joint architecture that combines mobility actions and power actions. For the single PI architecture shown in Fig. 5 (a), we observe that all mobile-APs move towards static UE within their neighborhood to improve system performance through mobility. However, this architecture only considers the mobility of the agent itself to maximize the reward value, resulting in the neglect of some UEs in the system, such as UE 2, UE 4, and UE 8, and the inability to provide uniform service quality. For the joint PI and PE architecture, compared to Fig. 5 (a), the joint mobility and power scheme undoubtedly provides a better moving trajectory as it avoids multiple mobile-APs serving the same static UE and fully considers each UE within the system. This is because power action can effectively regulate the interference among agents during the movement process, to encourage all agents to move in the direction with less interference, including adjusting the movement direction of mobile-AP 3 from UE 7 with higher interference to UE 4 with lower interference.

To further show the advantage of the joint PI and PE architecture, Fig. 6 and Fig. 7 show the cumulative distribution function (CDF) of achievable sum SE over different optimization schemes for the two architectures investigated. We notice that the joint PI and PE architecture undoubtedly achieves higher sum SE performance than that of the single PI architecture, as the joint architecture with collaborative power control mechanism is more competitive in suppressing interference between agents. Moreover, the proposed joint architecture has good adaptability and is beneficial for improving SE performance in various MARL-based schemes, e.g., 8.2\%, 15.31\%, and 15.59\% sum SE improvement for SF-MADDPG, Pe-MADDPG, and AIRMN, respectively, compared to the single PI architecture. More importantly, the proposed SF-MADDPG is effective in improving the sum SE performance due to its combination of GNN-aided collaborative architecture, permutation architecture, and directional decoupling architecture. For example, compared with MADDPG, AIRMN, Co-MADDPG, and Pe-MADDPG, the proposed SF-MADDPG yields a 45.97\%, 25.29\%, 18.79\%, and 14.41\% improvement in sum SE under a joint scalable framework. Meanwhile, we can also observe that the proposed algorithm outperforms conventional heuristic-based fraction power control schemes in SE performance by 81.63\%.
Besides, note that the proposed SF-MADDPG may perform worse than conventional MADDPG in severe interference situations, due to the initial exploration space being compressed by permutation architecture.

Moreover, Fig. 8 compares the proposed scalable framework with centralized MDGNN in terms of the CDF of achievable sum SE. We can notice that although centralized MDGNN fully utilizes global information, its inability to achieve mobility for channel reconstruction results in system performance far inferior to our proposed SF-MADDPG even under MDGNN (3D) with high complexity and low information loss. For example, compared with MDGNN (1D), MDGNN (2D), and MDGNN (3D), our proposed SF-MADDPG improves system performance by 21.87\%, 25.18\%, and 31.81\%, respectively. This indicates that effective channel reconstruction is crucial for breaking through system performance bottlenecks.

Fig. 9 shows the sum SE as a function of the number of mobile-APs with different schemes investigated under the single PI architecture, including LIIR and AIRMN based on directional decoupling architecture, Co-MADDPG based on GNN-aided collaborative architecture, Pe-MADDPG based on permutation architecture, and the proposed SF-MADDPG. We find that the sum SE for all optimization schemes grows with the number of mobile-APs $M$, with SF-MADDPG, Pe-MADDPG, and Co-MADDPG schemes showing relatively fast growth trends. Moreover, similar to the results obtained in Fig. 6 and Fig. 7, the proposed SF-MADDPG combines GNN-aided collaborative architecture, permutation architecture, and directional decoupling architecture to facilitate higher sum SE performance. For instance, the SE performance gap between the proposed SF-MADDPG and conventional Pe-MADDPG and Co-MADDPG under the single PI architecture fluctuates around 10\%, which is not affected by the increase in the number of mobile-APs and has good scalability.
\begin{figure}[t]
\centering
    \vspace{-0.3cm}
    \includegraphics[scale=0.5]{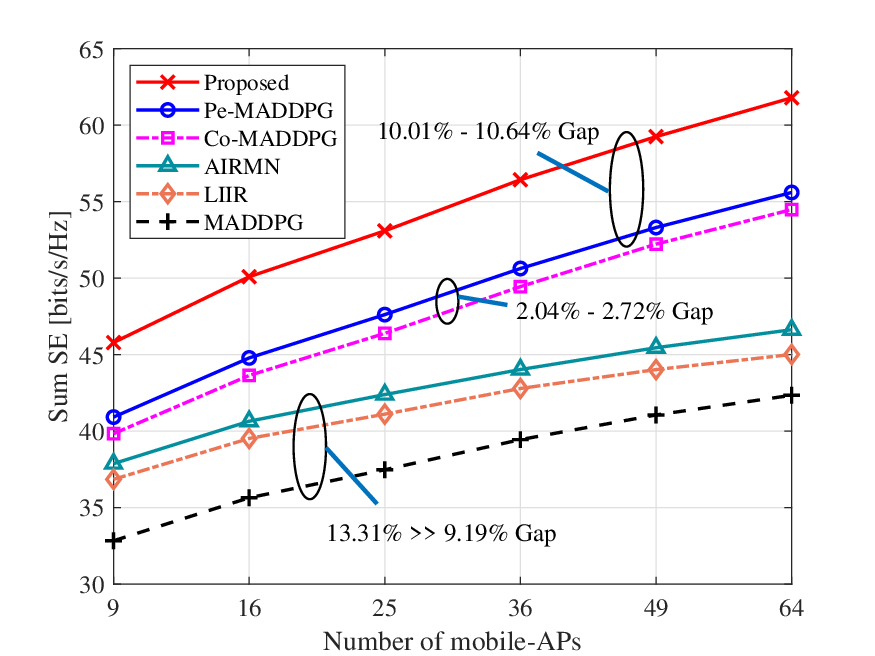}
    \vspace{-0.1cm}
    \caption{Sum SE against the number of mobile-APs $M$ under a single PI architecture over different MARL-based optimization schemes with $K=8$, $N=8$, $\tau_p=K$, and $\Delta = \lambda/2$.
    \label{fig1}}
\end{figure}
\begin{figure}[t]
\centering
    \includegraphics[scale=0.5]{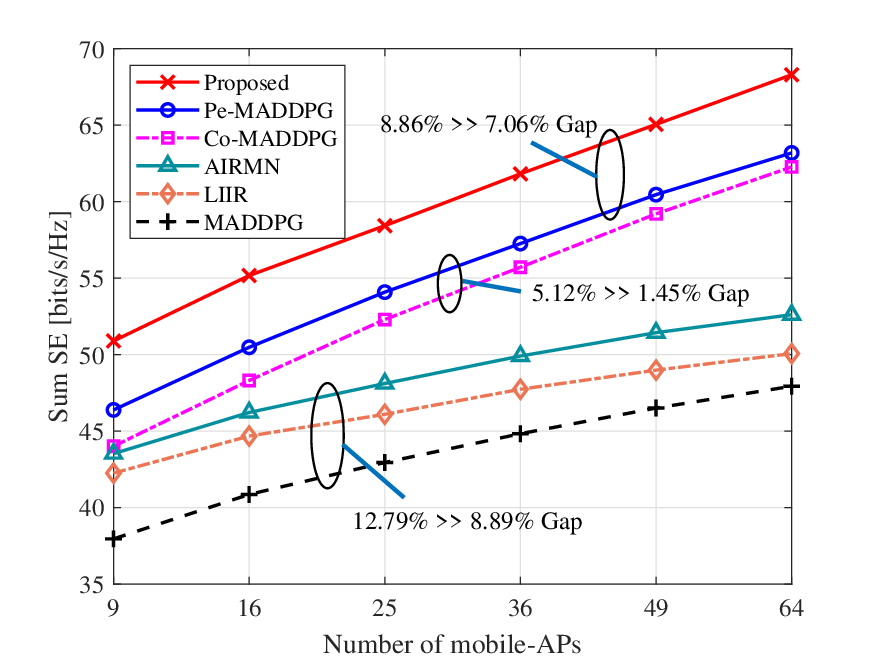}
    \vspace{-0.1cm}
    \caption{Sum SE against the number of mobile-APs $M$ under the joint PI and PE architecture over different MARL-based optimization schemes with $K=8$, $N=8$, $\tau_p=K$, and $\Delta = \lambda/2$.
    \label{fig1}}
\end{figure}
\begin{figure}[t]
\centering
    \vspace{-0.3cm}
    \includegraphics[scale=0.5]{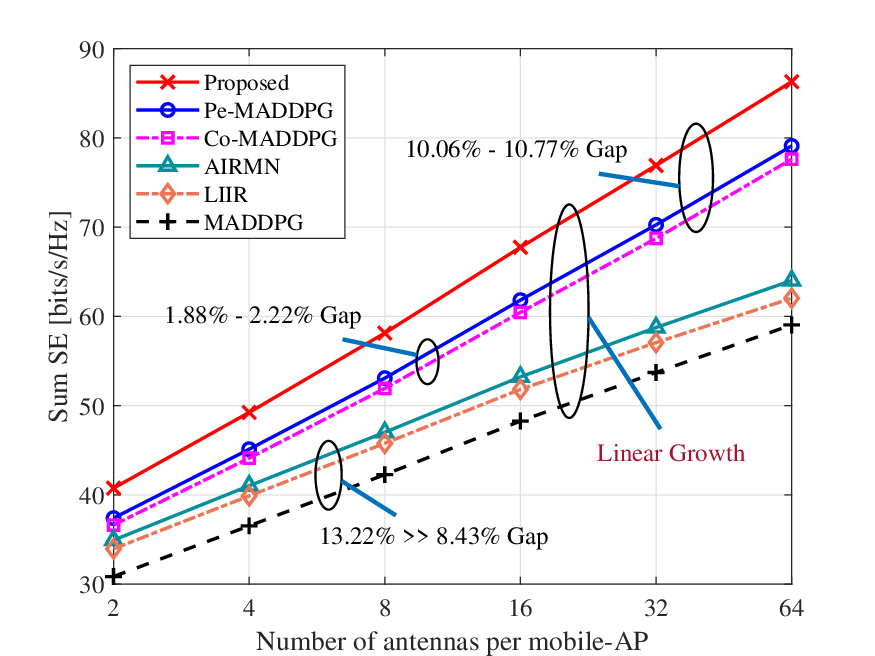}
    \vspace{-0.1cm}
    \caption{Sum SE against the number of antennas per mobile-AP $N$ under the joint PI and PE architecture over different MARL-based  optimization schemes with $M=16$, $K=9$, $\tau_p=K$, and $\Delta = \lambda/2$.
    \label{fig1}}
\end{figure}
\begin{figure}[t]
\centering
    \includegraphics[scale=0.5]{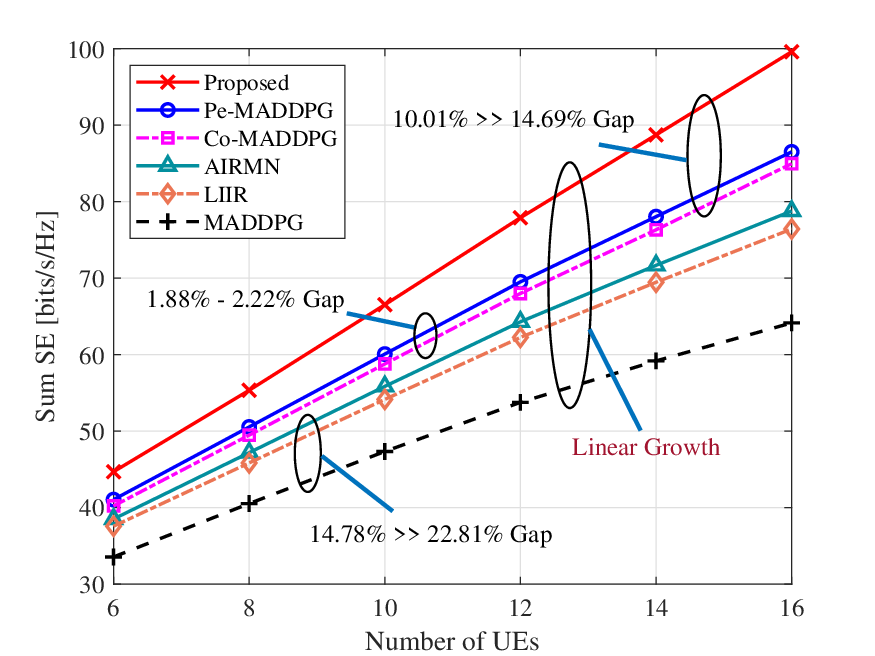}
    \vspace{-0.1cm}
    \caption{Sum SE against the number of UEs $K$ under the joint PI and PE architecture over different MARL-based optimization schemes with $M=25$, $N=8$, $\tau_p=K$, and $\Delta = \lambda/2$.
    \label{fig1}}
\end{figure}

To demonstrate the applicability of the proposed algorithms under different architectures, Fig. 10 discusses the sum SE as a function of the number of mobile-APs with different optimization schemes under the joint PI and PE architecture. We notice that, for all optimization schemes investigated, their SE performance still increases with N the number of mobile-APs $M$, which implies that the adaptability of the proposed algorithm is not affected by the type of permutation action, including mobility actions that are not sensitive to the order of entities and power actions that are sensitive to the order of entities. Moreover, we observe that compared to a single PI architecture, the SE performance growth rate of SF-MADDPG under the joint architecture may be weakened, e.g., from the original 10.64\% to 7.06\%. This indicates that actions in PI architectures that are not sensitive to the order of entities are more advantageous for improving SE performance.
\begin{table*}[t]
\centering
\fontsize{9}{9.5}\selectfont
\caption{Comparison of Details Between the Proposed SF-MADDPG and MADDPG.}
\label{paper}
\begin{tabular}{ccc}
\toprule
\bf Algorithms & \bf Proposed SF-MADDPG  & \bf MADDPG  \\
\midrule
\bf Deployed Networks (Centralized) & An extrinsic Critic network & $L$ Critic networks  \\
\bf Computational Complexity & $\mathcal{O}\left(\sum_{h=1}^{H_\mathrm{ec}}L(NK+L)Q_h^2\right)$  & $\mathcal{O}\left(\sum_{h=1}^{H_\mathrm{c}}L^2(NK+L)Q_h^2\right)$   \\
\midrule
\bf Deployed Networks (Decentralized) & $L$ DHPN and $L$ mixed Critic networks & $L$ Actor networks  \\
\bf Computational Complexity & $\mathcal{O}\left(\left(\sum_{h=1}^{H_\mathrm{d}}Q_h^2+\sum_{h=1}^{H_\mathrm{mc}}Q_h^2\right)L(NK+L)NK\right)$  & $\mathcal{O}\left(\sum_{h=1}^{H_\mathrm{a}}L(NK+L)NKQ_h^2\right)$   \\
\midrule
\bf Observation Dimension  &  $\mathcal{O}(L+(NK)K!)$ & $\mathcal{O}\left((K+L)L!+(NK)K!\right)$ \\
\midrule
\bf Convergence Rate (Total 5000 Episodes)   & 1607 \bf(-43.53\%) &  2846 \\
\bottomrule
\end{tabular}
\end{table*}
\begin{figure}[t]
\centering
    \vspace{-0.3cm}
    \includegraphics[scale=0.5]{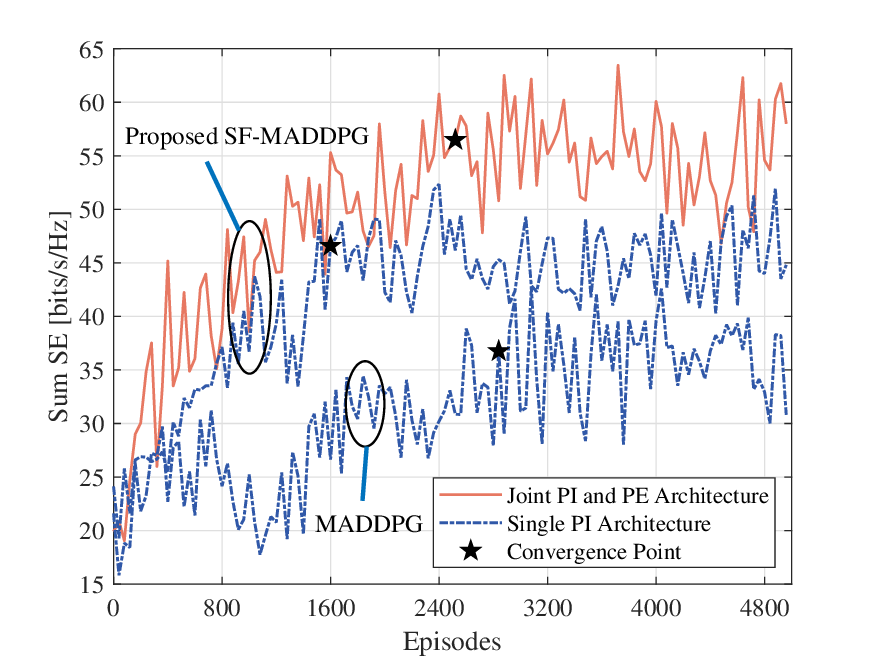}
    \vspace{-0.1cm}
    \caption{Convergence examples of the proposed SF-MADDPG and MADDPG with $M=9$, $K=8$, $N=8$, $\tau_p=K$, and $\Delta = \lambda/2$.
    \label{fig1}}
\end{figure}

Moreover, to further show the degree of influence of different types of permutation actions, Fig. 11 and Fig. 12 investigate the sum SE as a function of the number of antennas per mobile-AP $N$ and the number of UEs. As observed, the SE performance of all optimization schemes can still increase with $N$ or $K$, and the growth trend is not as weakened as in Fig. 9 and Fig. 10, always showing a linear growth trend.
More importantly, Fig. 12 reveals that the SE performance gap between the proposed SF-MADDPG and conventional schemes increases with the number of UEs, e.g., from 10.01\% to 14.69\%, which is different from the results presented when the parameters at the AP change. This demonstrates that directional decoupling architecture is more effective in large numbers of UEs, optimizing SE performance by dividing reward contributions for all agents, which is consistent with the advantages presented by LIIR and AIRMN. For example, compared to MADDPG, the SE performance gap has shifted from a decrease to an increase.

\begin{figure}[t]
\centering
    \vspace{-0.3cm}
    \includegraphics[scale=0.5]{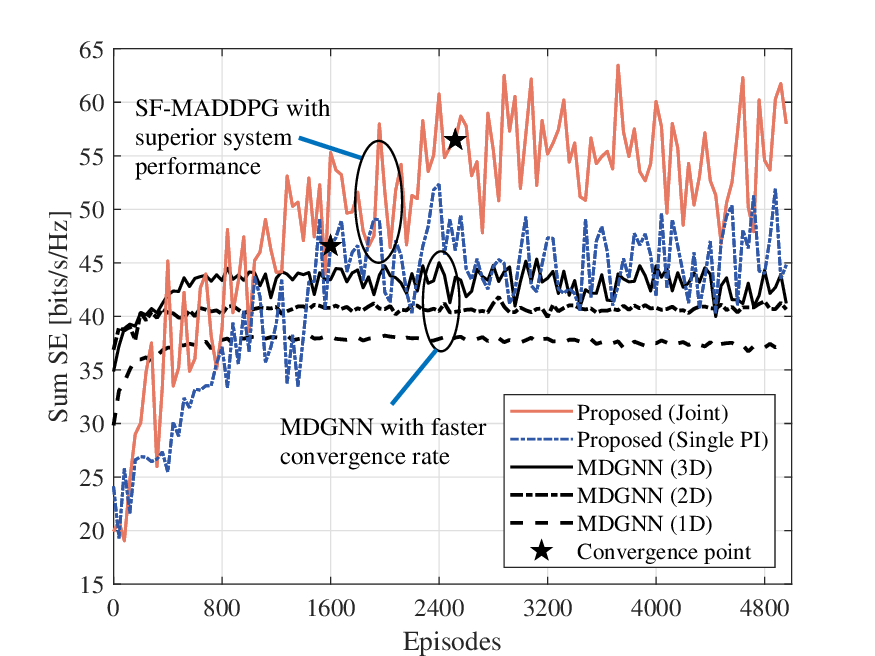}
    \vspace{-0.1cm}
    \caption{Convergence examples of the proposed SF-MADDPG and MDGNN with $M=9$, $K=8$, $N=8$, $\tau_p=K$, and $\Delta = \lambda/2$.
    \label{fig1}}
\end{figure}
Fig. 13 illustrates the comparison of the convergence rate over different optimization schemes, where the convergence points of the three training curves (i.e., in order from top to bottom as shown in Fig. 13) are 2523, 1607, and 2846. As observed, the proposed SF-MADDPG algorithm achieved a faster convergence rate compared to the original MADDPG algorithm, which increased by 43.53\%.
This is because the SF-MADDPG scheme utilizes a permutation network to permute the sorting of observation information, which is conducive to appropriately compressing the observation dimension of actions under PI properties (e.g., mobility actions that are not sensitive to the order of entities), reducing it to the original $1/L!$ times. Moreover, the joint architecture requires a higher number of episodes to achieve convergence compared to a single architecture, which is since actions under PE properties (e.g., power actions that are sensitive to the order of entities) need to correspond one-to-one with observations, resulting in the inability to compress the corresponding observation dimension. Meanwhile, building upon the exceptional convergence rate demonstrated by the proposed SF-MADDPG algorithm as analyzed earlier, enhancing real-time interaction capabilities can be achieved through the integration of advanced optimization techniques such as parallel computing, model compression, transfer learning, and more. This integrated approach aims to better align with the stringent demands of real-time communication and high reliability in practical scenarios.

Fig. 14 shows the comparison of the convergence rate between the proposed SF-MADDPG and centralized MDGNN, including MDGNN (1D) that only replaces one dimension, as well as MDGNN (2D) and MDGNN (3D) that replace joint dimensions. We can clearly observe that compared with the centralized MDGNN algorithm that utilizes global information, although the proposed SF-MADDPG algorithm has certain disadvantages in convergence rate, its full utilization of mobility to achieve channel reconstruction results in much better system performance than centralized MDGNN. This further illustrates the importance of utilizing mobility to achieve channel reconstruction for improving system performance.

Furthermore, Table \uppercase\expandafter{\romannumeral3} presents a comparison of different optimization schemes, where $H_\mathrm{ec}$, $H_\mathrm{d}$, and $H_\mathrm{mc}$ represent the hidden layers of the the extrinsic critic network, DHPN, and mixed critic network under the proposed SF-MADDPG, while $H_\mathrm{a}$ and $H_\mathrm{c}$ represent the hidden layers of the critic network, DHPN and actor network under the conventional MADDPG.
We note that the proposed SF-MADDPG can optimize the order insensitive related actions of entities by adopting PI properties, reducing its observation dimension from the original $\mathcal{O}\left((K+L)L!+(NK)K!\right)$ to $\mathcal{O}(L+(NK)K!)$. This further reveals that utilizing permutation properties to distinguish the types of actions, i.e., entity-uncorrelated actions and entity-correlated actions, is beneficial for compressing observation dimension and improving the convergence rate.
Although SF-MADDPG increases the computational complexity of decentralized networks by deploying additional mixed critic networks, it only requires a centralized external critic network for policy updates, making the overall computational complexity of SF-MADDPG comparable to conventional MADDPG. This indicates that the proposed scalable framework still meets the deployment requirements of practical scenarios.
\section{Conclusion}
In this paper, we initially investigated the downlink performance of a cell-free mMIMO system equipped with mobile-APs and considered a joint optimization problem of mobility and power control to effectively enhance coverage and suppress interference. Then, we proposed a novel scalable framework with MARL to address the challenges of high computational complexity, poor collaboration, limited scalability, and uneven reward distribution, fully unleashing the potential of downlink cell-free mMIMO systems. In numerical results, we investigated the impact of the number of mobile-APs, UEs, and antennas per mobile-AP, and compared the SE performance for different optimization schemes and network architectures. It is greatly important to find that using permutation properties to distinguish between entity-uncorrelated and entity-correlated actions is beneficial for compressing the observation space and improving the convergence rate.
In future work, we will investigate the downlink power control with both mobile-APs and UEs equipped with multiple antennas and explore the performance impact of multi-antenna UEs on the effectiveness of our proposed framework.
\bibliographystyle{IEEEtran}
\bibliography{IEEEabrv,Ref}
\end{document}